\documentclass[12pt,preprint]{aastex}

\shorttitle{Properties of Dwarf Galaxies, Local Sample}
\shortauthors{O. Vaduvescu, M. L. McCall, M. G. Richer}
\slugcomment{Accepted in AJ (for August 2007 issue)}

\begin{document}

\title{Chemical Properties of Star Forming Dwarf Galaxies}

\author{Ovidiu Vaduvescu\altaffilmark{1}}
\affil{
Instituto de Astronom\'{i}a, Universidad Cat\'{o}lica del Norte \\
       Avenida Angamos 0610, Antofagasta, Chile \\
Former address: ACRU \& SAAO, University of KwaZulu-Natal, \\
       School of Mathematical Sciences, Durban 4041, South Africa }

\author{Marshall L. McCall\altaffilmark{2}}
\affil{
York University, Department of Physics and Astronomy \\
       4700 Keele Street, M3J~1P3, Toronto, ON, Canada}

\author{Michael G. Richer\altaffilmark{3}}
\affil{
Observatorio Astr\'{o}nomico Nacional, Instituto de Astronomia \\
       Universidad Nacional Aut\'{o}noma de M\'{e}xico \\
       PO Box 439027, San Diego, CA 92143-9027, USA}

\altaffiltext{1}{email: ovidiuv@yahoo.com}
\altaffiltext{2}{email: mccall@yorku.ca}
\altaffiltext{3}{email: richer@astrosen.unam.mx}

\begin{abstract}

Recent studies of the near-infrared (NIR) properties of dwarf irregular galaxies (dIs) 
and blue compact dwarfs (BCDs) have provided improved estimates for the NIR luminosity 
of old stellar populations in these galaxies. These can be used to derive gas fractions, 
and thereby to evaluate how BCDs have evolved with respect to dIs. Oxygen abundances
have been derived for four BCDs in the Virgo Cluster from a run at Gemini-North in 2003.
Combining these new abundances with published values, we study the correlations among 
the metallicity, $K_s$ luminosity, gas mass, baryonic mass, and gas fraction. Within 
errors, the two types of dwarfs appear to share a common relation between the oxygen 
abundance and the luminosity in $K_s$. The correlation between metallicity and the gas 
fraction is the same for BCDs as for dIs, indicating that BCD evolution has been similar 
to dIs. Since dIs appear to have evolved as isolated systems, the BCD bursts are 
unlikely to be a consequence of gas infall or merging. 

\end{abstract}


\section{Introduction}

\subsection{A Universe of Dwafs}

Dwarf galaxies are by far the most numerous galaxies in the Universe. About 80--90 per
cent of the members of the Local Group are classified as dwarfs \citep{mat98,gre00}, while
about 85 per cent of the known galaxies in the Local Volume ($D\la10$ Mpc) are also dwarfs
\citep{kar04}. The space density of dwarfs in the Universe has been suggested to be
about 40 times that of bright galaxies \citep{sta92}.

{\it Dwarf galaxies} are defined arbitrarily as galaxies having an absolute magnitude
fainter than $M_B \sim -16$ mag \citep{tam94} or $M_V \sim -18$ mag \citep{gre00,gre01}.
Based on their optical appearance, dwarf galaxies can be classified into five groups: 
dwarf irregulars (dIs), blue compact dwarfs (BCDs), dwarf ellipticals (dEs), dwarf 
spheroidals (dSphs), and dwarf spirals (dS's). The distinction between dEs and dSphs 
is not clear (e.g., \citealp{bin94}), while dS's can be regarded as the very small end 
of spirals \citep{mat97,gre01}. Excluding dS's, dIs and BCDs are the only systems 
harboring active or recent star formation activity. According to hierarchical clustering 
models of galaxy formation, larger galactic structures build up and grow through the
accretion of dwarf galaxies \citep{whi91,kau93}. 
Nevertheless, some recent evidence (e.g., \citealp{ven04,hel06}) suggests that the halo 
of the Milky Way could not have been constructed from dSphs as we presently know them. 
In this picture, dIs and BCDs may be similar to the building blocks for more massive 
galaxies. As such, they are important probes for studying matter in its near-primordial 
state, thus being relevant to understanding galaxy formation and evolution. 

Despite the advances in the last decades, the relation between dIs and BCDs remains
unclear. Some authors suggest that BCDs and dIs are the same type of galaxy, with BCDs
being dIs undergoing bursts of star formation (e.g. \citealp{thu85}). Others argue that
dIs are a fundamentally distinct type from BCDs, with no simple evolutionary links between
them (e.g., \citealp{jam94}). \citet{ric95} argued on chemical grounds that BCDs may be 
more akin to dSphs than dIs. \citet{pap96a} review previous arguments concerning the 
evolutionary relationships between BCDs, dIs and dEs without reaching firm conclusions.

\subsection{Determining Chemical Abundances for Dwarfs}
\label{abund_dwarfs}

The {\it metallicity}, $Z$, is defined as the fraction of elements other than hydrogen and
helium by mass (e.g., \citealp{pag97}). Metallicity is a key parameter intimately linked to
the formation and evolution of galaxies. It gives an indication of age in the sense that it
reveals how far the conversion of gas into stars has proceeded. In practice, the metallicity
in star-forming galaxies is quantified via the {\it oxygen abundance}, defined as the 
fraction of oxygen by number, $12+$log($n$(O)/$n$(H)). 
After hydrogen and helium, the most abundant element in the Universe is oxygen. It is 
convenient to measure the oxygen abundance because it exhibits emission lines in HII region 
spectra which are very bright and easy to measure. 

Physical and chemical properties of nebulae can be derived by measuring ratios of
intensities of collisionally-excited emission lines (e.g., \citealp{ost89}). Recombination 
lines of metals are normally too faint to detect in extragalactic nebulae, though in some 
of the very brightest objects recombination lines have been detected (e.g., \citealp{pei06}). 
The radiation emitted by a nebula depends upon the abundances of the chemical elements 
and the physical state of the gas, especially its average temperature and density. 

A natural reference for elemental abundances is the Sun, which has an oxygen abundance,
$12+\log(\rm{O/H})_{\odot}=8.69$ (e.g., \citealp{all01}). 
We define Z$_{\odot}$ = (O/H)$_{\odot}$. 
It is customary to define ``metal poor'' as a system with an abundance lower 
than this value, and ``metal-rich'' as a system with an abundance exceeding this value. 
dIs and BCDs are the most metal-poor galaxies, having metallicities as low as 
$\sim1/50$ Z$_{\odot}$. In high-redshift clouds of gas, which may be building blocks 
of today's galaxies, the metallicity can reach values as low as 0.001 Z$_{\odot}$ 
\citep{kun00}. Stars in our Galaxy reach metallicities as low as 0.00001 Z$_{\odot}$ 
\citep{spa00}. 

There are two commonly-used methods to determine oxygen abundances in HII regions. 
The {\it direct method} ($T_e$) is founded upon a direct measurement of temperature. 
It is applicable whenever the temperature diagnostic [O~III]$\lambda4363$ is detectable and 
in which the doubly ionized O$^{+2}$ ion represents the dominant form of oxygen \citep{ost89}.
The most accurate oxygen abundances are derived using the direct method, but for most 
galaxies [O~III]$\lambda4363$ is faint and difficult to detect. 
When [O~III]$\lambda4363$ is not detected, 
the {\it bright-line method} ($R_{23}$), originally proposed by \citet{pag79}, 
is commonly used.
The oxygen abundance is parametrized as a function of the ratio 
$R_{23} = (I\rm{[OII]}\lambda3727 + I\rm{[OIII]}\lambda\lambda4959,5007)/I(H\beta)$. 
Unfortunately, the $R_{23}$ indicator is not a monotonic function of oxygen abundance; 
for a given value of $R_{23}$, two values of the oxygen abundance are possible. 
However, \citet{mcg91} suggested that $\log([\rm{NII}]\lambda6583/[\rm{OII}]\lambda3727)$ 
can be used as a discriminator between the lower and upper branches. 
To estimate oxygen abundances, we adopt the calibration of $R_{23}$ developed by 
\citet{mcg94}, which employs $O_{32}=I(\rm{[OIII]}\lambda4959,5007)/I(\rm{[OII]}\lambda3727)$ 
to discriminate between the two branches. 

\subsection{Dwarfs and the Closed Box Model}

One of the simplest frameworks for galaxy evolution is the {\it closed box model} 
\citep{sea72,pag97}.
According to this model, a galaxy consists initially of gas with no stars and no metals.
The stellar initial mass function (IMF) is assumed to be constant in time. Stars that 
end their life as supernovae are assumed to enrich the interstellar gas with metals 
immediately, thereby eliminating time as a variable. Throughout the entire time, the 
galaxy evolves as a closed system, with no mass inflow or outflow. For such a system,
the metallicity at any given time is solely determined by the fraction of baryons which
remains in gaseous form, which is referred to as the {\it gas fraction}.

Many researchers, starting with \citet{leq79}, have examined how the metallicity of 
a galaxy depends upon its mass or luminosity, finding that more massive galaxies are 
more abundant in metals. 
\citet{ski89} confirmed a strong correlation between absolute magnitude and metallicity 
for local dIs, in the sense that the least luminous dIs are the most metal-poor systems. 
\citet{ric95} derived a more robust $L_B-Z$ relationship, based on a sample of 25 nearby 
irregular galaxies having well-determined and self-consistent distances along with abundances
calculated from measured [O~III] temperatures.
An enlarged sample was considered by \citet{mil96}, with the same conclusions. The same 
correlation was found by \citet{van97} in another sample of 15 gas-rich, low surface 
brightness dwarfs. 

Information about the chemical evolution of dwarf galaxies is buried within the $L-Z$ 
relation, but it is extremely difficult to extract.  This is because the relation is 
affected by both flows (which may influence the metallicity, but not the luminosity) and 
the detailed history of star formation (which affects noth metallicity and luminosity). 
Luminosity is a proxy for stellar mass, but how representative of stellar mass it is 
depends upon the bandpass. 

In a series of four papers, \citet{mel02}, \citet{mel04}, \citet{leej04} and \citet{sal05} 
studied the $L-Z$ relationship of giant ``starburst emission-line galaxies'' (ELGs), using 
data in the visible and NIR. The authors derived a linear relationship with a steeper slope 
than previously found for dwarfs, finding that the slope decreases as the wavelength of the 
luminosity bandpass increases. 
Using spectra of 6387 emission-line star-forming galaxies, \citet{lam04} 
confirmed a linear $L-Z$ relation in the local Universe ($z<0.15$) over a large range of 
abundances ($\sim2$ dex) and luminosities ($\sim9$ mag). \citet{tre04} utilized SDSS 
imaging and spectroscopy of about 53,000 star-forming galaxies at $z\sim0.1$, finding 
linear $L-Z$ relationships, although contours appear to level off at high luminosities 
($M_g<-20$). 

A more useful relation for studying chemical evolution is the correlation between 
metallicity and the gas fraction. \citet{lee03a} and \citet{lee03b} used data in the 
visible to show that the oxygen abundance in dIs is tightly correlated with the gas 
fraction. The result was used to argue that dIs have evolved in isolation, without 
inflow or outflow of gas. 
The observations pointed to a subsolar yield (e.g., \citealp{gar02}), but the 
significance of this is difficult to assess because values for yields are critically 
dependent upon methods used to determine abundances. 
However, yields are relative, being critically dependent on methods used to determine 
abundances. 
A somewhat different conclusion was reached by \citet{van06b}, who measured abundances 
in 21 apparently isolated dIs selected based on their morphological classification and 
distances determined mostly from systemic velocities. While several galaxies in their 
sample follow the closed box model, they argue that either outflow of enriched gas or
inflow of pristine gas has occurred in most galaxies. 

\subsection{NIR Imaging of Dwarfs}

The utility of light as a gauge of galaxy mass depends upon the wavelength of observation.
In star--forming galaxies such as dIs and most BCDs, the young population shines brightly
in the visible, overwhelming the light from the old stellar component, which traces the bulk
of the mass. Therefore, observations in the visible do not necessarily reflect the 
stellar mass of a galaxy. This problem can be reduced in the near--infrared (NIR), where
the intermediate--age and old populations become more visible. Furthermore, light in the NIR
is much less attenuated by extinction (absorption of light due to internal and Galactic dust)
than light in the visible.

\citet{vad05a} (hereafter referred to as Paper I) and \citet{vad06} (hereafter referred to
as Paper II) studied the structural properties of dIs and BCDs in the NIR. The dI sample
(Paper I) included 28 dwarfs from the Local Volume closer than ~5 Mpc having accurate 
distances from the literature (i.e., derived using Cepheids and the tip of the red giant 
branch). The BCD sample (Paper II) included 16 dwarfs in the Virgo Cluster, all 
approximated to have the same distance modulus ($DM=30.62$). The surface brightness profiles 
of dIs were fitted using a sech law, which models the light at all radii, while BCDs were
fitted using a sech function to model the diffuse old component plus a Gaussian to model 
the young starburst (at small radii). Absolute sech magnitudes were derived using the known 
distances and correcting for the extinction. These data are used here to gauge the stellar 
masses of the systems and thereby make possible a detailed study of chemical evolution.

\subsection{The Present Paper}

In the present paper, we analyse the chemical properties of our dI and BCD samples. 
We use the NIR data gathered in Papers I and II, in conjunction with metallicities taken
from the literature and derived from our own spectra, to derive fundamental relations 
relevant to the formation and evolution of dwarfs and to compare dIs and BCDs chemically. 

In Section~\ref{observations}, we present spectral observations of four Virgo BCDs
which we observed at Gemini North, and in Section~\ref{reduction} their reduction. 
Sections~\ref{chemical} and \ref{mass-met} address chemical properties of dIs and BCDs, 
considering correlations among luminosity, metallicity, mass (in various forms), and 
the gas fraction. In Section~\ref{evolution}, we approach the implications for evolution, 
comparing dIs and BCDs in the context of the closed box model. Section~\ref{conclusions} 
presents our conclusions.

\section{Gemini Observations}
\label{observations}
Using Gemini-North with GMOS, we detected [O~III]$\lambda4363$ in two of the four 
BCDs which we observed. 
Long-slit spectra of four BCDs in the Virgo Cluster (VCC~24, VCC~459, VCC~641, and 
VCC~2033) were acquired in queue mode on two nights between Feb 28 and Mar 2, 2003. 
The data were taken with the Gemini Multi-Object
Spectrograph (GMOS) mounted at the f/15.82 focus of the Gemini-North 8.1~m telescope located
atop Mauna Kea, Hawaii. The slit width was 0.5\arcsec~and the slit length was 5.5\arcmin.
The GMOS camera includes three CCDs with $2048\times4608$ pixels each, creating a mosaic of 
$6144\times4608$ pixels. Pixel sizes are 13.5 $\rm{\mu m}$, so the scale is 0.0824\arcsec/pixel
along the dispersion axis and 0.0727\arcsec/pixel along the slit axis. We used the B600\_G5303
grating which gave a spectral resolution of 0.45 \AA/pixel. Data were acquired both in the blue
(central wavelength of the grating 4680 \AA, range 3298-6062 \AA) and in the red (central
wavelength 5600 \AA, range 4218-6982 \AA). Data were binned $4\times4$. Air masses were
between 1.0 and 1.2, so the slit was not oriented along the parallactic angle. 

Dome flat fields were acquired every night for all instrumental settings, interspersed 
between blue and red galaxy spectra, and also between blue and red standard star spectra. 
Biases were taken at the end of the run. One spectroscopic standard star, Feige 66, was 
observed once every night. We include the log of our observations in Table~\ref{tbl-1}.

\section{Data Analyses}
\label{reduction}

\subsection{Reductions}

Images were reduced using the GEMINI v.1.6 GMOS package under IRAF using the sequence 
given in \citet{vad05b}. Bias and flat field frames (the latter taken for each instrument 
setting) were applied first, then individual spectra were combined for each grating setting. 
The arc lamp spectra were corrected for bias and flat field and used to calibrate the combined 
red and blue spectra. The two-dimensional spectra were collapsed to one dimension. The same 
sequence was applied to the standard star spectra. Spectra were calibrated using the standard 
star observations, correcting for atmospheric extinction. The blue and red calibrated spectra 
were joined into final spectra, using the common lines H$\beta$~$\lambda$4861, 
[O~III]$\lambda$4959 and [O~III]$\lambda$5007 to equalize flux scales. The scale factors 
ranged between 1.3 and 2.4. 

\begin{figure}
\epsscale{1.0}
\plotone{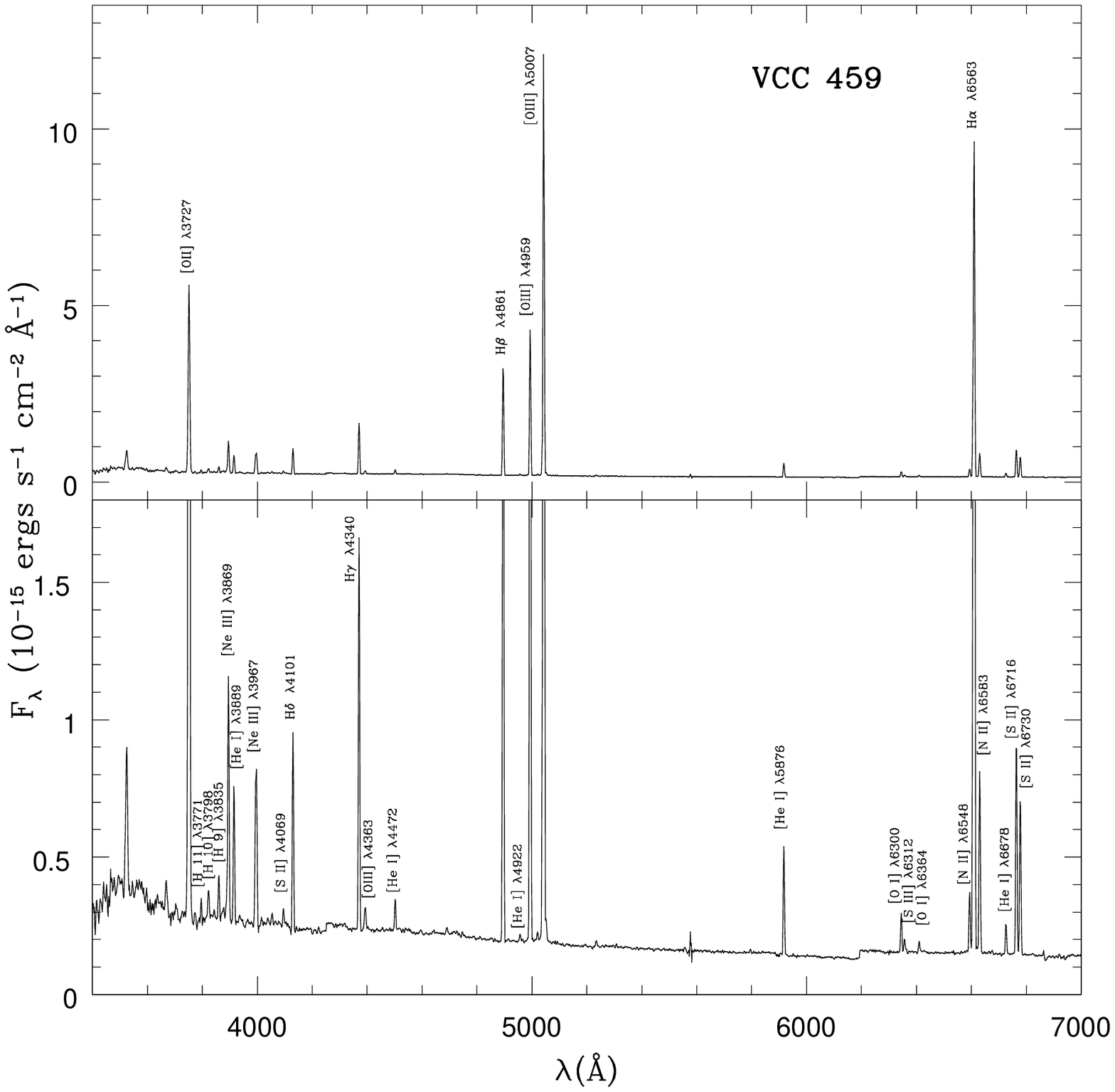}
\caption{
\label{sp_VCC459comb}
Reduced combined spectrum of VCC 459.
For a better view of the lines, the spectra are shown at two vertical scalings. 
A ``fracture'' due to imperfect combination of the continua is visible at 6200 \AA.
This did not affect measurements of the emission lines.
}
\end{figure}

\begin{figure}
\epsscale{1.0}
\plotone{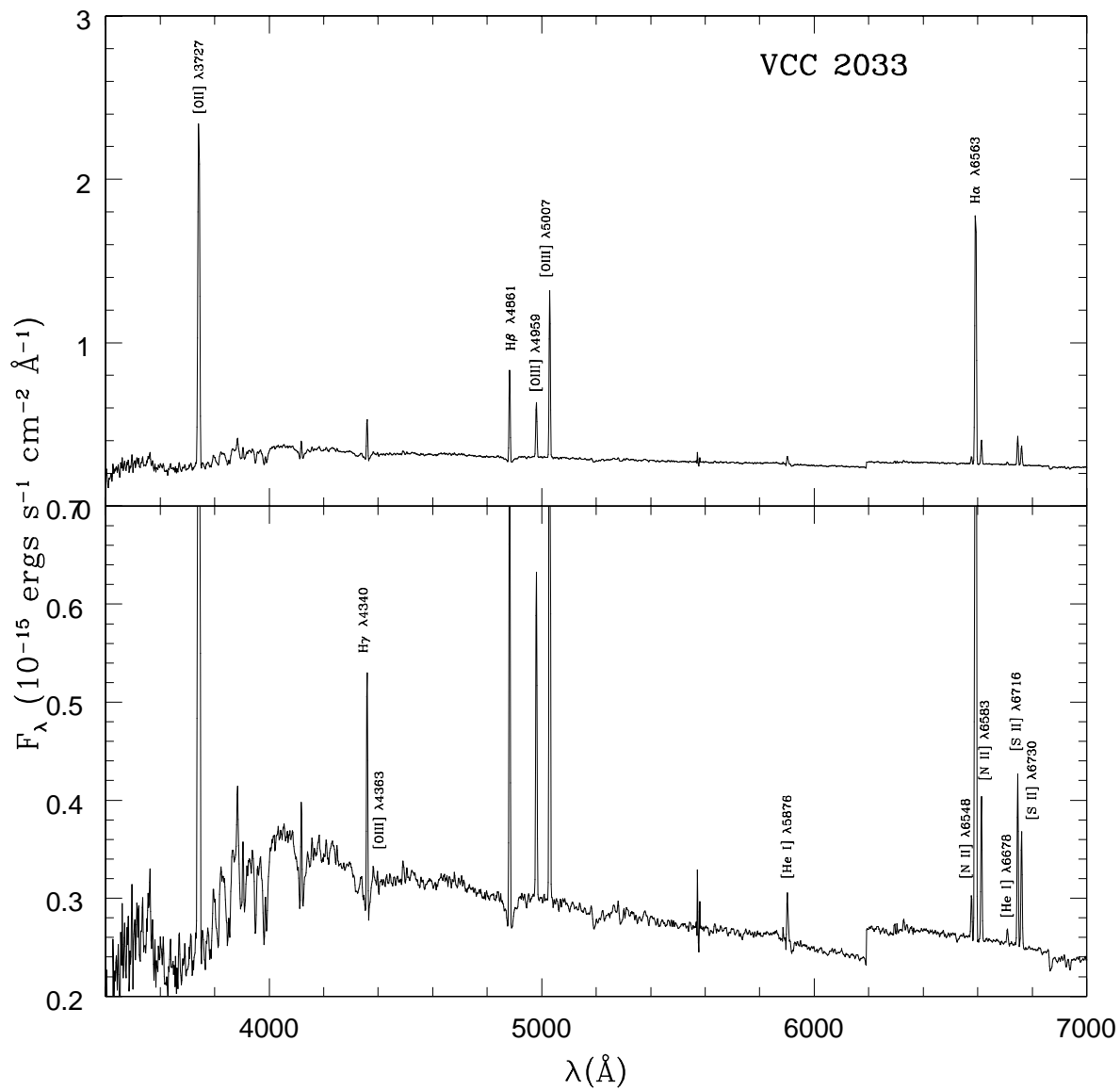}
\caption{
\label{sp_VCC2033comb}
Reduced combined spectrum of VCC 2033.
For a better view of the lines, the spectra are shown at two vertical scalings. 
A ``fracture'' due to imperfect combination of the continua is visible at 6200 \AA.
This did not affect measurements of the emission lines.
}
\end{figure}

\begin{figure}
\epsscale{1.0}
\plotone{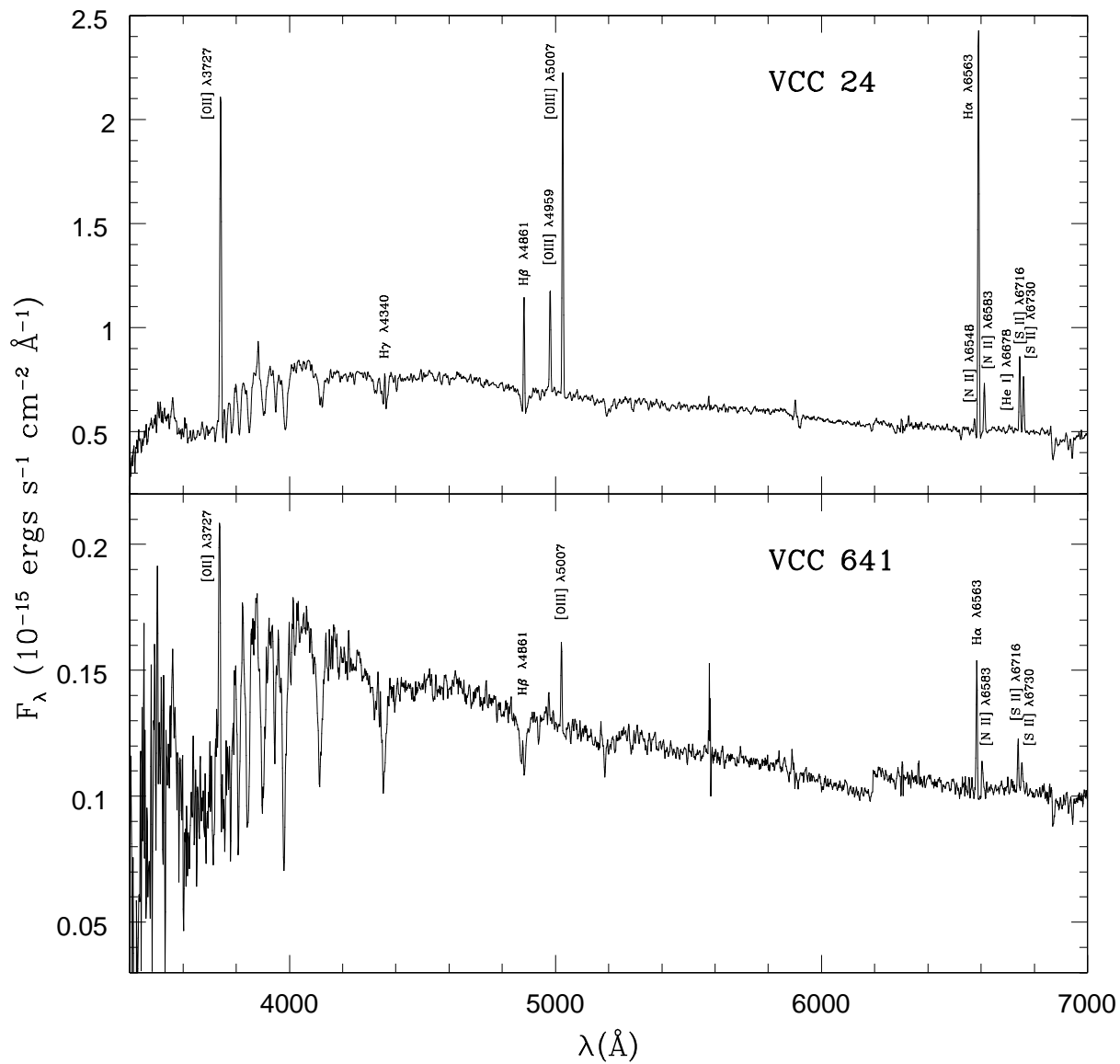}
\caption{
\label{sp_VCC24641comb}
Reduced combined spectra of VCC 24 and VCC 641.
A ``fracture'' due to imperfect combination of the continua of VCC 641 is visible at 6200 \AA.
This did not affect measurements of the emission lines.
}
\end{figure}

Figures~\ref{sp_VCC459comb} to \ref{sp_VCC24641comb} present the reduced combined 
spectra of VCC~459, VCC~2033, VCC~24, and VCC~641. The first two are presented using
different scalings to show the bright and faint lines. The last two show fewer lines, 
so they are displayed using a single scaling. In a few cases, a ``fracture'' due to 
imperfect combination of the red and blue spectra is visible in the overlapping region 
at 6200 \AA. Because the scaling of the red and blue spectra was based entirely upon 
emission line fluxes, measurements of the emission lines should be unaffected. 

\subsection{Line Measurements}

We detected [O~III]$\lambda4363$ in the core of two Virgo BCDs, namely VCC~459 and VCC~2033, for 
which it was possible to determine the oxygen abundances using the direct method. In VCC~24 and 
VCC~641 we did not detect [O~III]$\lambda4363$, but did measure stronger lines. For them, we 
were able to use the bright-line method to derive abundances.

Emission lines were measured using INTENS, a FORTRAN program developed by \citet{mcc85}. 
This is a non-interactive non-linear least-squares fitting program 
which can efficiently fit and measure emission and/or absorption lines in one-dimensional 
spectra (flux versus wavelength) or two-dimensional spectra (flux versus wavelength as a 
function of spatial position). All spectra analyzed in this paper were one-dimensional. 

\begin{figure}
\epsscale{1.0}
\plottwo{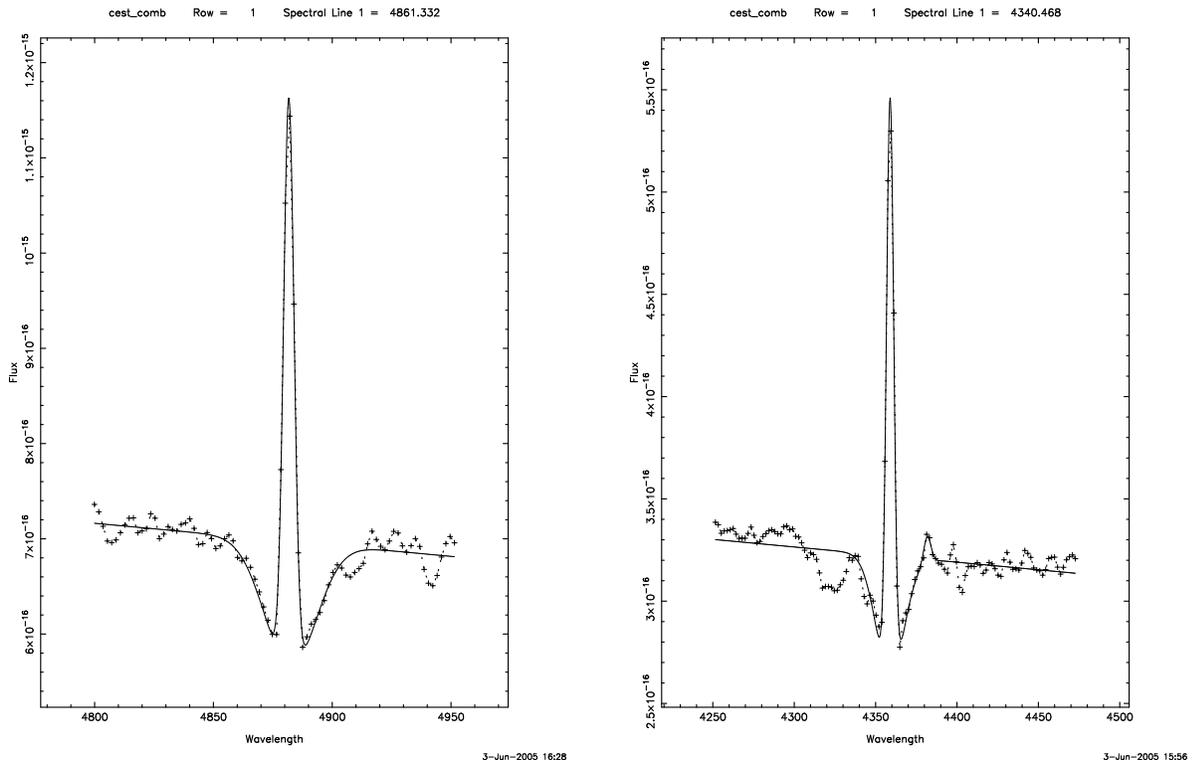}{f4b.eps}
\caption{
\label{INTENS_FITS}
Flux versus wavelength for two INTENS fits. Left: H$\beta$ in VCC~24;
Right: H$\gamma$ and [OIII]$\lambda4363$ in VCC~2033. Pluses mark observations, and solid curves are fits.
}
\end{figure}

A few of the hydrogen lines exhibited obvious underlying Balmer absorption, especially
at H$\beta$. Using INTENS, a simultaneous fit of emission and absorption profiles allowed for
a direct determination of the flux in the emission and the equivalent width of the underlying
Balmer absorption. Measured emission fluxes are given in Table~\ref{VCC_lines}. 
In Figure~\ref{INTENS_FITS}, we illustrate two fits. 

\subsection{Abundances}

We used SNAP (Spreadsheet Nebular Analysis Package; \citealp{kra97}) to derive
abundances from ratios of line fluxes. SNAP is an add-in for Microsoft Excel$^{\rm{TM}}$. 
From appropriately chosen spectral diagnostics, SNAP will compute reddening, apply
equivalent width and reddening corrections, compute electron temperatures and densities,
and compute ionic abundances.

Using SNAP, the extinction was calculated using $F(H\alpha)/F(H\beta)$ ratio, assuming
the reddening model of \citet{fit99} and that the dust is located in the Galactic foreground. 
The redshift of each galaxy was taken from NED. If the [O~III]$\lambda4363$ line was detected, 
then the temperature was calculated using $I($[O~III]$\lambda5007)/I($[O~III]$\lambda4363)$. 
Otherwise, the temperature was assumed to be $T_e=10,000$ K. The electron density was computed 
using the $I([\rm{S~II}]\lambda6716)/I([\rm{S~II}]\lambda6731)$ ratio. 
If the [S~II] lines were poorly measured, a density of 100 electrons/cm$^3$ was assumed. 

Once the data were corrected for reddening, the ionic abundances of O$^+$ and O$^{++}$ 
relative to H$^+$ were calculated for those objects for which [O~III]$\lambda4363$ was 
detected. These were then summed to get $12+$log($n$(O)/$n$(H)). 

We include in Table~\ref{VCC_lines} the reddening-corrected line fluxes with respect 
to H$\beta$ for VCC~24, VCC~459, and VCC~2033. For VCC~641, H$\beta$ was severely affected 
by underlying absorption, so we report its line fluxes with respect to H$\alpha$ instead. 

Table~\ref{table_BCDs_props} lists the measured chemical properties of the four BCDs observed.
For VCC~2033, [O~III]$\lambda4363$ was marginally detected, and thus, the oxygen abundance error
is large (0.3 dex). Using the bright line $R_{23}$ method \citep{mcg91,mcg94}, we 
derive $12+\log(\rm{O/H})=8.31$, which is 0.3 dex higher than 
the [O~III]$\lambda4363$ value ($12+\log(\rm{O/H})=7.99 \pm 0.30$). 
The $R_{23}$ value seems to agree better with the measurement of \citet{vil03}, who give lower
and upper limits of 8.11 and 8.40, respectively. Due to the large error in our
[O~III]$\lambda4363$ measurement, we adopt for VCC~2033 an oxygen abundance which is the average 
value of the [O~III]$\lambda4363$ and the $R_{23}$ measurements, i.e., $12+\log(\rm{O/H})=8.15 \pm 0.15$.
For VCC~641, H$\beta$ was not detected, so in this case we used the emissivity calculations to 
determine the ratio H$\alpha/$H$\beta$, then we employed the bright line method. 
\citet{van06a} compare the reliabilty of empirical and semi-empirical methods to derive 
abundances, suggesting that the typical error in an abundance derived from $R_{23}$ is around 
0.2 dex. We include this error for all our $R_{23}$ results in Tables~\ref{table_BCDs_props}, 
\ref{dIs_met}, and \ref{BCDs_met}.

\section{Chemical Properties of Dwarf Galaxies}
\label{chemical}

\subsection{Our Luminosity-Metallicity Relation for dIs}

In Table~\ref{dIs_met}, we include the oxygen abundances for 17 dwarfs classified as dIs from 
our Local Volume sample, out of 27 observed in the NIR (Paper I), together with their sources. 
In the second part of the table we include four dwarfs from Virgo observed by us in the NIR 
(Paper II), classified as dIs based on their luminosity profiles. 
In the last part of the Table we append another 8 dIs with known metallicities from the literature 
for which $K_s$ luminosities are determined from 2MASS data using the best distances available 
in the literature (Fingerhut 2005, personal communication). Note that 2MASS stars were used 
to derive zero-points for our observed fields (Papers I and II), so the photometric scales are 
consistent. Nevertheless, one must be wary of 2MASS magnitudes for galaxies, owing to the 
short exposure times and the correspondingly bright detection-thresholds. 

\begin{figure}
\epsscale{1.0}
\plotone{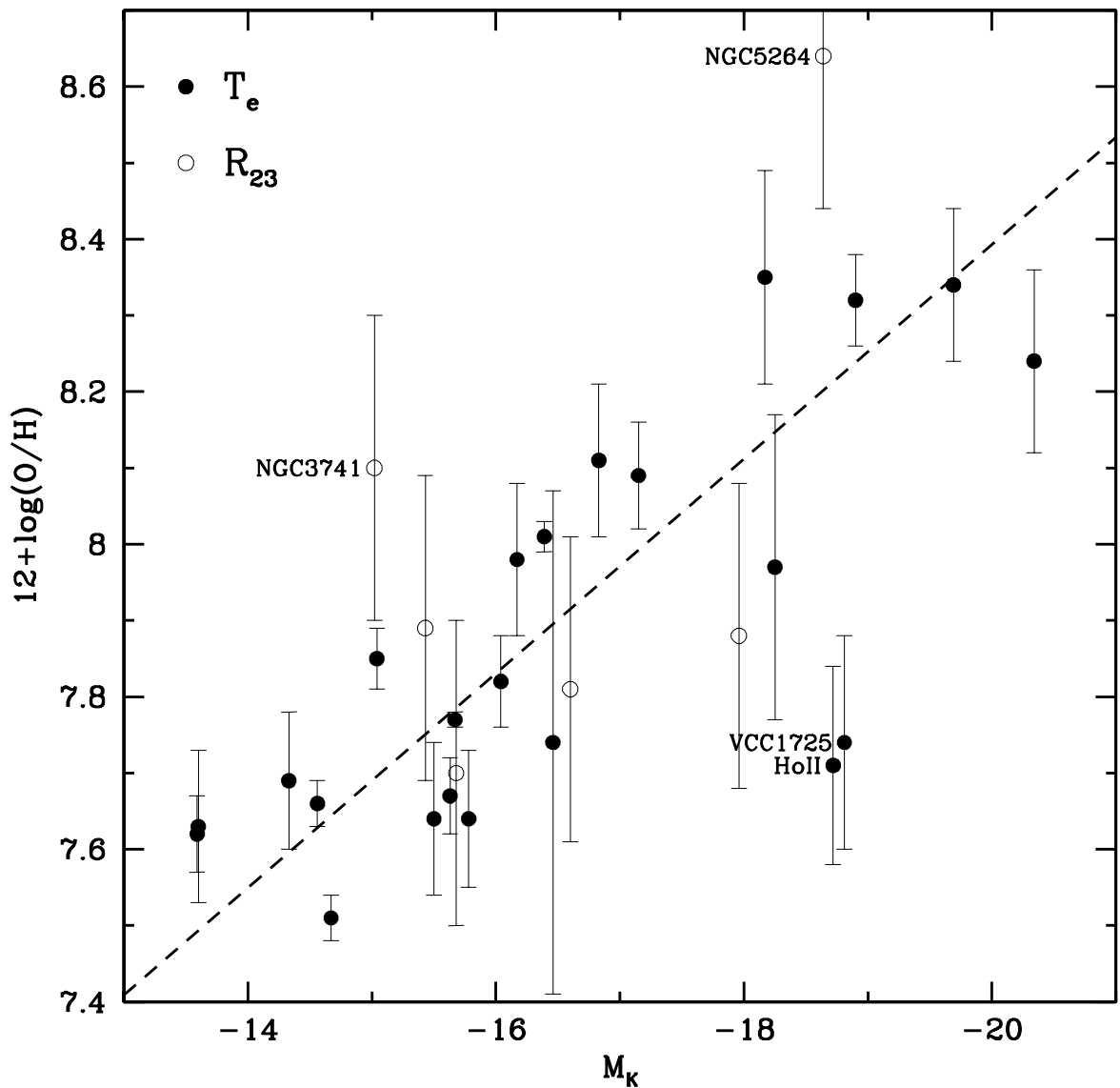}
\caption{
\label{LZ_fig_dIs}
The oxygen abundance versus absolute sech magnitude in $K_s$ for dIs. 
$T_e$-based abundances are plotted as solid symbols, and $R_{23}$-based abundances as open symbols. 
Errors in metallicities are shown as vertical bars. Typical errors for absolute magnitudes are $\sim0.1$ 
mag. Four outliers are labeled in the plot. The geometric mean fit to other dIs is plotted as a dashed line. 
}
\end{figure}

In Figure~\ref{LZ_fig_dIs}, we plot the $L_K-Z$ relationship for the 29 dIs in Table~\ref{dIs_met}. 
For the galaxies observed by us, we use sech absolute magnitudes, while for the ones observed by 
2MASS, we use total absolute magnitudes. 
$T_e$-based abundances are plotted as solid symbols, and $R_{23}$-based abundances as open symbols. 
We include errors in metallicities; typical errors for absolute magnitudes are about 0.1 mag. 
There is an excellent correlation. Four outliers appear in the plot: NGC~5264 and NGC~3741, 
whose metallicities are derived using the bright line method, Holmberg II, whose $K_s$ magnitude 
comes from 2MASS, and VCC~1725. 

Following \citet{lee03a}, a best-fit line for the correlation between two parameters with
comparable errors can be obtained with the geometric mean functional relationship \citep{dra98},
which assumes similar dispersions in both observables. An equal weighting of points is assigned
for both variables in a given fit. The geometric mean functional relationship relies upon the
minimization of the sum of areas bounded by the shortest horizontal and vertical lines from
each data point to the best-fit line.

Excluding the four labeled points, we obtain the following L$_K$-Z relationship for 25 dIs 
from Table~\ref{dIs_met}: 
\begin{equation}
12 + \log(\rm{O/H}) = (-0.141 \pm 0.015) M_K + (5.581 \pm 0.244)
\label{LZ_eq_dIs}
\end{equation}

\noindent
We plot the fit as a dashed line in Figure~\ref{LZ_fig_dIs}. Most of the points are
consistent with the fit, given the errors in metallicities. A nearly identical
relationship is obtained in $J$: 
\begin{equation}
12 + \log(\rm{O/H}) = (-0.140 \pm 0.014) M_J + (5.692 \pm 0.214)
\end{equation}
\noindent
The rms dispersion in abundances in the $L-Z$ relations in $K_s$ and $J$ is 0.11 and 0.10, 
respectively. Similar relations are obtained using the isophotal and total absolute magnitudes 
instead of the sech magnitudes. 

\subsection{Our Luminosity-Metallicity Relation for BCDs}

In Table~\ref{BCDs_met}, we include 15 dwarfs classified as BCDs based on the literature and
luminosity profiles. In the first part of the table, we include 12 BCDs from Virgo (Paper II),
and, in the last part, the three dwarfs from the Local Volume classified as BCDs based on
their luminosity profiles (Paper II) and \citet{van00}. 

\begin{figure}
\epsscale{1.0}
\plotone{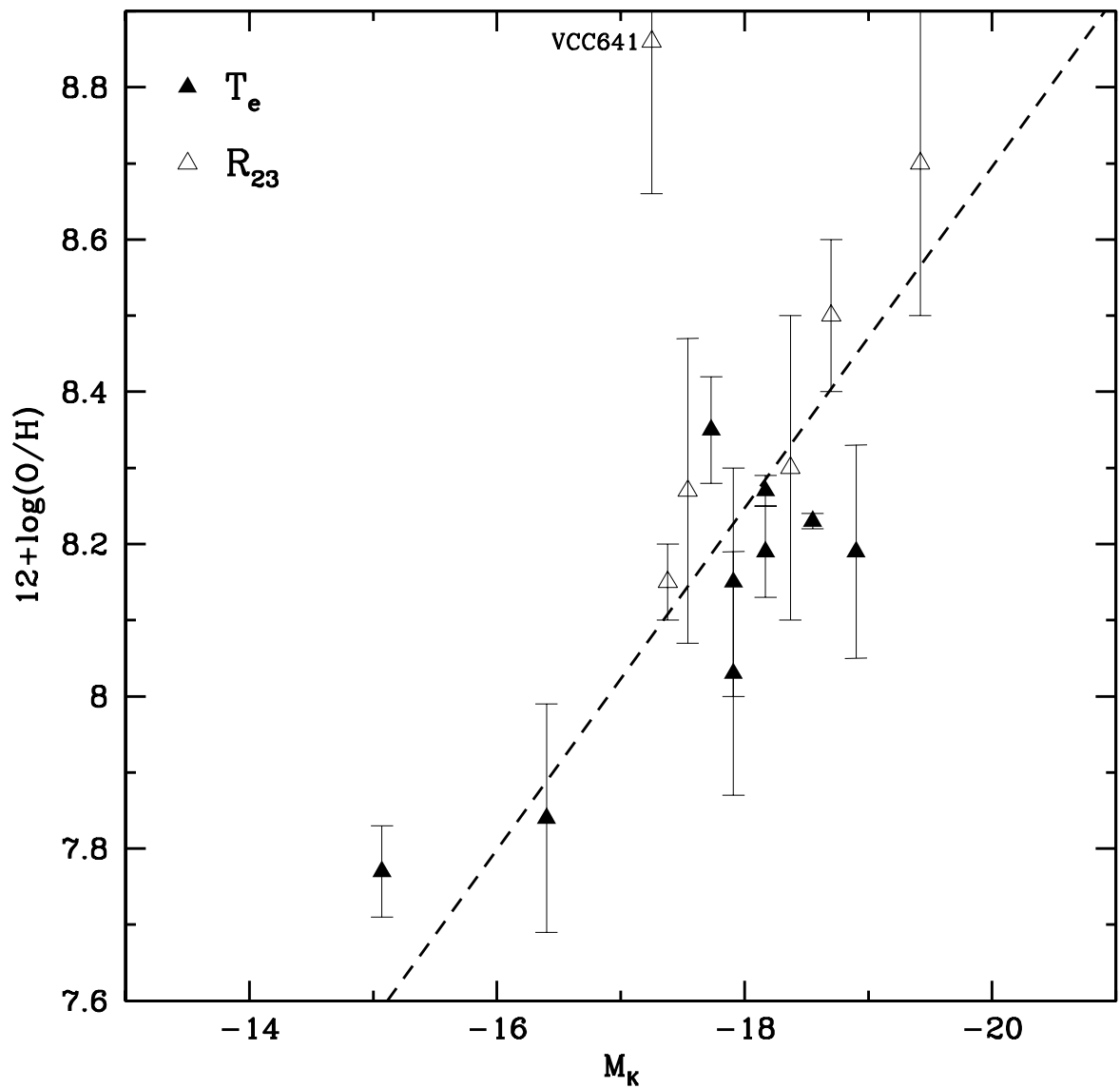}
\caption{
\label{LZ_fig_BCDs}
The oxygen abundance versus absolute sech magnitude in $K_s$ for BCDs. 
$T_e$-based abundances are plotted as solid symbols, and $R_{23}$-based abundances as open symbols. 
We include errors in metallicities.
Typical errors for absolute magnitudes are $\sim0.1$ mag. One outlier is labeled. The geometric 
mean fit to the other BCDs is plotted as a dashed line. 
}
\end{figure}

In Figure~\ref{LZ_fig_BCDs}, we plot the $L_K-Z$ relationship for 15 BCDs from Table~\ref{BCDs_met}. 
To sample the underlying stellar population, we employed sech magnitudes. 
$T_e$-based abundances are plotted as solid points, and $R_{23}$-based abundances as open triangles.
We include errors in metallicities; typical errors for absolute magnitudes are about $0.1$ mag.
One outlier appears in the plot, VCC~641, having a very uncertain oxygen abundance.
Excluding this point, we obtain the following $L_K-Z$ relation for 14 BCDs from geometric mean 
fitting: 
\begin{equation}
12 + \log(\rm{O/H}) = (-0.224 \pm 0.030) M_K + (4.212 \pm 0.537)
\label{LZ_eq_BCDs}
\end{equation}
\noindent
We plot the fit as a dashed line in Figure~\ref{LZ_fig_BCDs}. A similar $L_J-Z$relationship is
obtained in $J$:
\begin{equation}
12 + \log(\rm{O/H}) = (-0.216 \pm 0.028) M_J + (4.530 \pm 0.473)
\end{equation}
\noindent
The rms dispersion in abundance in the $L-Z$ relations in $K$ and $J$ is 0.12 and 0.11, 
respectively. Similar relations are obtained using the isophotal and total magnitudes 
instead of the sech magnitudes. 

\begin{figure}
\epsscale{1.0}
\plotone{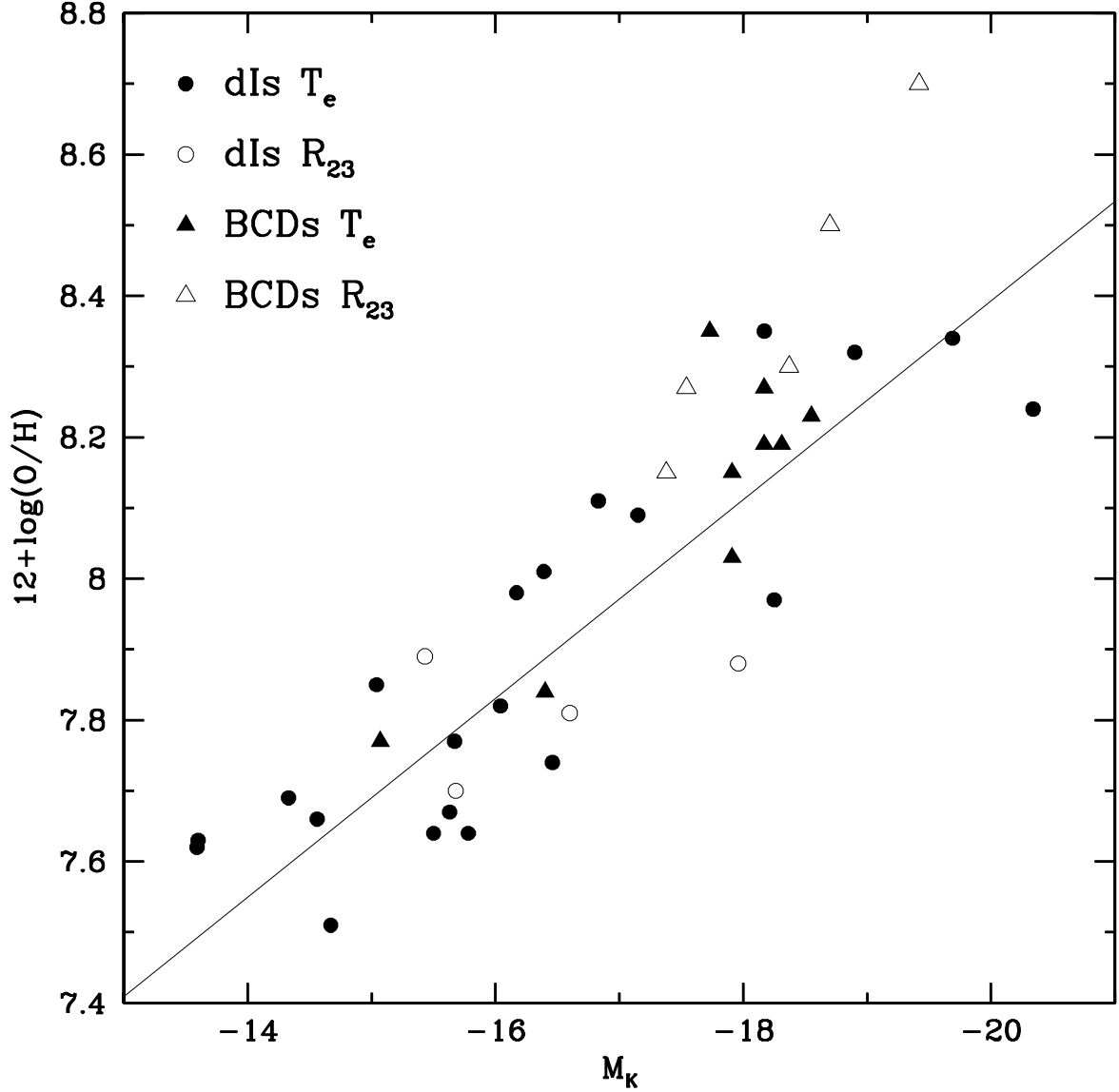}
\caption{
\label{LZ_fig_dIs_BCDs}
The oxygen abundance versus absolute sech magnitudes in $K_s$ for dIs and BCDs. 
dIs are plotted as circles, while BCDs as triangles. 
$T_e$-based abundances are plotted as solid symbols, and $R_{23}$-based abundances as open symbols. 
The five outliers from Figures~\ref{LZ_fig_dIs}-\ref{LZ_fig_BCDs} are not shown. The solid line 
represents the $L-Z$ relation for the dIs. 
}
\end{figure}

In Figure~\ref{LZ_fig_dIs_BCDs}, we combine the dIs and BCDs in our samples. For visibility 
reasons, we do not include the five outlier points. The solid line represents the $L_K-Z$ 
relation for the dIs. Recall the sech magnitude for BCDs excludes the luminosity of the starburst 
and so represents the underlying stellar component. 
Figure~\ref{LZ_fig_dIs_BCDs} illustrates that, though the $L_K-Z$ relations for dIs and BCDs 
are formally different, the two samples largely overlap. 
Given the relatively poor luminosity distribution of our BCD sample ($M_K$ mostly between 
-18 and -20), more data are required to definitively decide whether the two $L_K-Z$ relations 
are really different. 

\subsection{Comparison with the Literature}

In Table~\ref{LZ_comp}, we summarize different $L-Z$ relationships presented in the literature. 
The $L-Z$ relation we derive here for local dIs is similar to those found by others in the $B$-band 
\citep{ski89,ric95,lee03a,van06b} and in the NIR \citep{sav05}. The coefficients derived by us in 
the NIR are very close to the ones derived by \citet{ric95} and \citet{van06b} in the visible 
employing mostly $T_e$-based abundances. Using NIR luminosities has not reduced the scatter 
about this relation for dIs, perhaps due to the difficulty of obtaining sufficiently deep NIR 
photometry. Recently, \citet{oli06} used data from Paper I and 2MASS to find a $L_K-Z$ relation 
very similar to ours but with larger dispersion. Their sample includes 29 dwarfs, of which six 
appear to be BCDs. \citet{lee06a} find less scatter in the $L-Z$ relation using mid-IR luminosities 
at 4.5 microns from the {\it Spitzer Space Telescope}. Assuming that most of the galaxies of 
\citet{sal05} are BCDs and that their total magnitudes approximate the old component, then their 
results are in agreement with ours, provided BCDs and dIs follow different $L-Z$ relations. 

The studies quoted in Table~\ref{LZ_comp} may be divided into local and more distant 
samples of dwarf galaxies. The slopes of the $L-Z$ relations for the local samples 
(\citealp{ski89,ric95,lee03a,shi05,oli06,lee06a}, and this work) are all very similar. 
There may be a difference between the slopes in the blue and 
infrared, but it is slight. The more distant samples of dwarfs (\citealp{lam04,tre04,sal05}),
however, all produce $L-Z$ relations with steeper slopes, reminiscent of our relation for 
BCDs. If these samples of more distant dwarfs are composed primarily of BCDs with some 
smaller fractions of dIs, the steeper slopes make sense. However, based upon our data, 
it is possible that the steeper slope arises from an imperfection in the calibration of 
$R_{23}$ at high metallicity. 

\section{Mass-Metallicity Relations}
\label{mass-met}

Gas-rich star forming dwarf galaxies are systems that are well mixed chemically, elemental 
abundances being observed to vary very little across their faces \citep{kob96,kob97,pag97}, 
having no signifiant spacial inhomogeneities in [O~III]$\lambda4363$ oxygen abundances measured 
from HII regions located at different galactocentric radii \citep{lee04,lee05,lee06b}. Chemical 
evolution can be explored via correlations between oxygen abundance and the stellar mass, gas 
mass, total mass, and the gas fraction.

Using data in $B$ for a sample of dIs observed in the field, \citet{lee03a} showed that the oxygen
abundance correlates with the stellar mass of the systems. However, in another sample of dIs in the
Virgo Cluster, the relation was more scattered. \citet{lee03b} found also that the relation with 
the gas mass was scattered for the field sample, and that some Virgo dIs were depleted in gas. 
The relation between metallicity and the gas fraction suggested that the evolution of some dIs in 
Virgo was affected by flows, such as stripping of gas due to the cluster neighborhood. 
The authors showed that this relationship could be used unequivocally to recognize gas 
deficiency in dwarfs. Below we address the relations between metallicity and mass, employing 
NIR photometry. 

\subsection{Gas Masses}

The total mass of gas can be gauged from the HI 21-cm flux, $F_{21}$, measured by radio 
observations (\citealp{huc00}; \citealp{huc01,huc03}). 
Data for dIs are summarized by \citet{kar04}. We include in Table~\ref{dIs_met} the 
logarithm of the total HI flux from the latter source. For Virgo BCDs, HI fluxes are given 
in Table~\ref{BCDs_met}. They are derived from \citet{gav05}, who synthesize a set of data 
for 355 late-type galaxies in the Virgo cluster with HI masses for many dwarfs derived from 
previous work by Hoffman and collaborators (\citealp{huc00}; \citealp{huc01,huc03}). 

The HI mass, $M_{HI}$, in solar masses is given by the following equation \citep{rob75,rob94}:
\begin{equation}
M_{HI} = 2.356 \times 10^5 F_{21} D^2
\label{HI_mass}
\end{equation}
\noindent
where $F_{21}$ is the 21-cm flux integral in Jy km s$^{-1}$ and $D$ is the distance in Mpc.
To account for helium and other metals, the total gas mass in solar masses is given by
\begin{equation}
M_{gas} = M_{HI}/X
\end{equation}
\noindent
where $X$ is the fraction of the gas in the form of hydrogen, here assumed to be 0.733 
\citep{lee03a}. 
In Table~\ref{dIs_met} we include the logarithm of the total gas mass for the dIs in our 
sample, calculated from $F_{21}$ and the distance modulus included in the first column. 
In Table~\ref{BCDs_met} we include the logarithm of the total gas mass for the BCDs in our 
sample. 

\begin{figure}
\epsscale{1.0}
\plotone{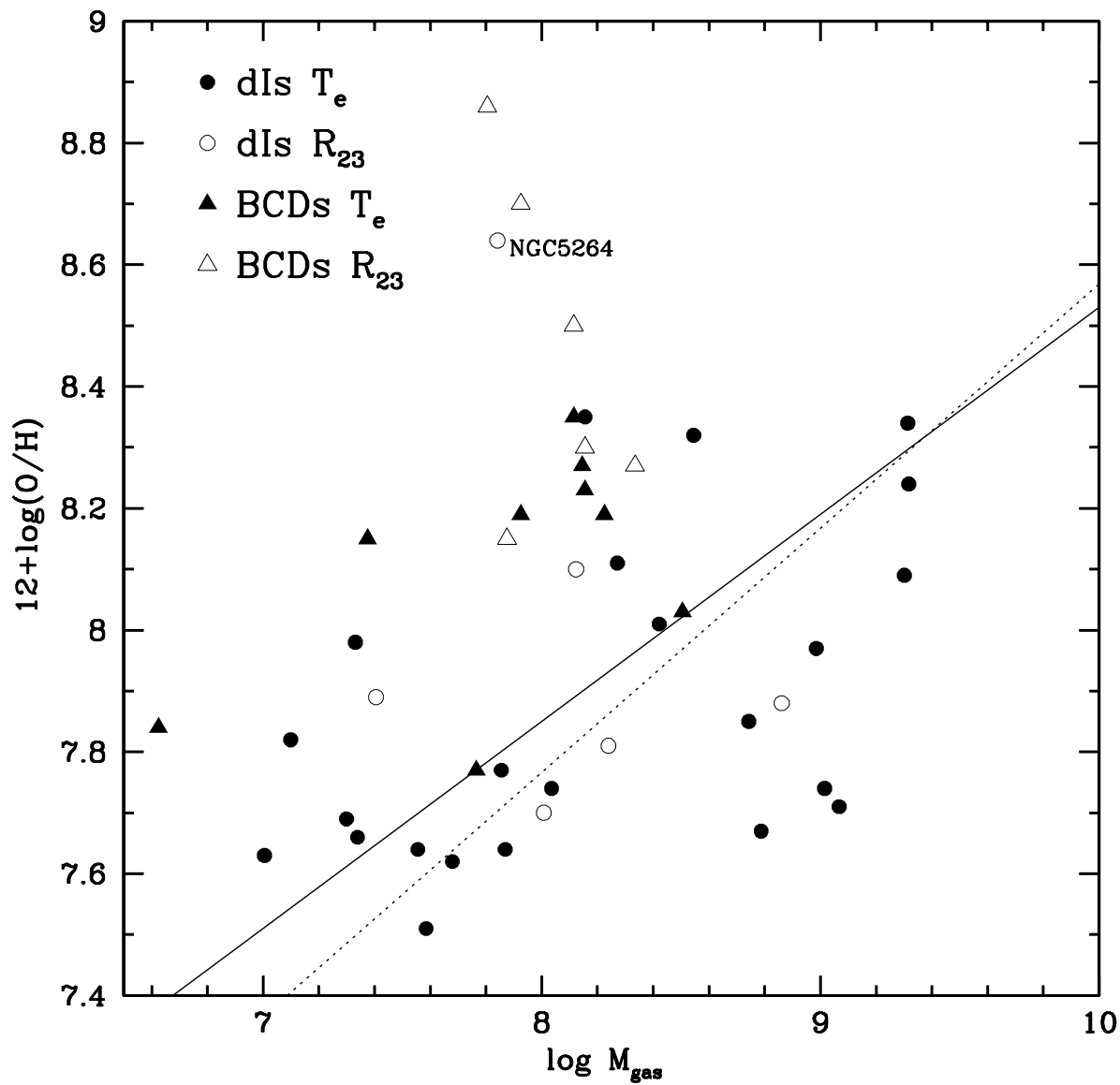}
\caption{
\label{met_gasmass}
Correlation between oxygen abundance and the gas mass. 
dIs are plotted with circles, and BCDs with triangles. 
$T_e$-based abundances are plotted as solid symbols, and $R_{23}$-based abundances as open symbols. 
dIs with a greater gas mass are more abundant in metals. 
Ignoring NGC~5264, the geometric linear fit to the dI data is marked by a solid line. For 
comparison, the fit found by \citet{lee03a} is shown with a dotted line. 
BCDs often have less gas than dIs at a given metallicity. 
}
\end{figure}

Figure~\ref{met_gasmass} plots the metallicity versus the logarithm of the total gas mass
(expressed in solar units). dIs are plotted with circles, and BCDs with triangles. Dwarfs 
having $T_e$ abundances are plotted with solid sybols and those having $R_{23}$ abundances 
with open symbols. The correlation for the dIs is very poor, with NGC~5264 again being an 
outlier in the sample. Rejecting this point, a geometric linear fit to the rest of the dI 
sample (28 objects) gives the following relation: 
\begin{equation}
12+\log(\rm{O/H}) = (5.13 \pm 0.45) + (0.34 \pm 0.06) \log(M_{gas})
\end{equation}
\noindent
This relation is very close to the fit found by \citet{lee03a}, who used data for 21 dIs. 
The two samples overlap, with the sample of \citet{lee03a} extending from $M_B=-18$ to $-11$ 
mag (corresponding to $M_K$ from $-21$ to $-14$), and our sample extending from $M_K=-20$ 
to $-13$. We plot in Figure~\ref{met_gasmass} our fit with a solid line and the fit of 
\citet{lee03a} with a dotted line. Some Virgo BCDs lie leftward of the dI envelope, i.e., 
less gas for a given oxygen abundance, a result found also in \citet{lee03b} for their Virgo 
dI sample, suggesting that they are depleted in gas relative to dIs of comparable metallicity. 
Either they are less massive systems overall, or gas has been removed, perhaps by dynamical 
interactions in the cluster environment. 

\subsection{Stellar Masses}

In Paper I, it was shown that the shape of the unresolved (diffuse) stellar component 
of nearby dIs in the NIR is approximated well by a hyperbolic secant (sech) function. 
In Paper II, it was demonstrated that the same function fits the shape of the diffuse 
component underlying star formation bursts in BCDs. It is believed that the luminosity 
associated with the sech model provides a good estimate of the contribution to the light
by stellar populations older than about 3 Gyr (Paper I). As such, it should be the best
gauge of total stellar mass. 

Expressed in solar masses, stellar masses can be calculated from
\begin{equation}
\log M_* = \log {\frac {M_*}{L_K}(old)} - 0.4 (M_K(old)-M_{K\sun})
\label{star_mass}
\end{equation}
\noindent
where $M_*/L_K(old)$ is the mass-to-light ratio of the old stellar component in $K_s$,
in solar units, $M_K(old)$ is the absolute magnitude of the old component in $K_s$, and 
$M_{K\sun}$ is the absolute magnitude of the Sun in $K_s$. Here, we prefer the sech 
magnitudes derived in Papers I and II as representative of the luminosity of the old 
population, $M_K(old)$. 
However, we augment the dI sample with galaxies detected by 2MASS for which catalogued
total magnitudes are used as estimates of $M_K(old)$. These galaxies are distinguished 
symbolically in plots. For the absolute magnitude of the Sun, we adopt $M_{K\sun}=+3.3$
\citep{bes98}. 

For low metallicity galaxies (0.004 Z$_{\odot}$ to 0.0004 Z$_{\odot}$), Portinari 
(2005, private communication) derived $M_*/L_K \sim 0.8\, M_{\sun}/L_{K\sun}$ using 
population synthesis models. \citet{dro04} obtained an identical result from composite 
stellar population models for ages between about 4 and 8 Gyr (see their Figure 1). From 
extensive modeling of the Milky Way, \citet{bis03} derived $M_*/L_K = 0.6$. 
Based on galaxy evolution models to investigate the relation between 
mass to light ratios and colours of spiral galaxies, \citet{bel03} predict  
$M/L_K\sim1\, M_{\sun}/L_{K\sun}$ for $B-R\sim2$ mag 
(which can be regarded as an average colour of dwarfs). 
Here, we adopt $M_*/L_K = 0.8\, M_{\sun}/L_{K\sun}$. 

\subsection{Baryonic Masses}

\begin{figure}
\epsscale{1.0}
\plotone{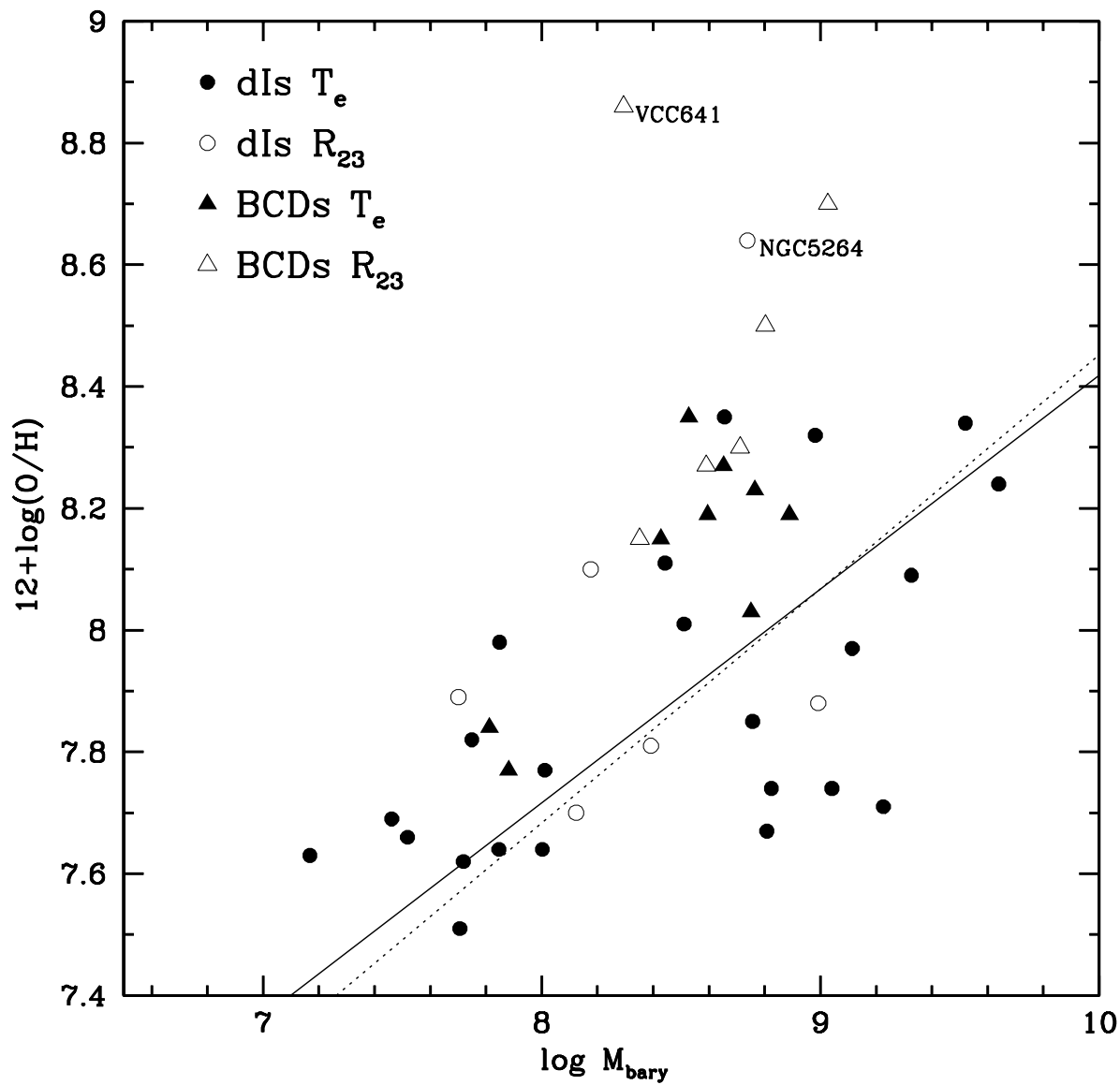}
\caption{
\label{met_totalmass}
Correlation between oxygen abundance and baryonic mass. dIs are plotted with circles, 
and BCDs with triangles. 
$T_e$-based abundances are plotted as solid symbols, and $R_{23}$-based abundances as open symbols. 
Stellar masses were derived from sech magnitudes in $K_s$. More massive dIs are more abundant in 
metals. Ignoring NGC~5264, the geometric linear fit to the dI data is drawn as a solid line. 
For comparison, the \citet{lee03a} fit is plotted with a dashed line. 
}
\end{figure}

Assuming baryonic matter consists of gas and stars only, the baryonic mass of a galaxy
can be expressed simply by the sum of the two components, $M_{bary} = M_{gas} + M_*$.
Figure~\ref{met_totalmass} plots the metallicity versus the logarithm of the baryonic
mass expressed in solar masses. dIs are plotted with circles, and BCDs with triangles. 
Dwarfs having $T_e$ abundances are plotted with solid symbols, while those having $R_{23}$ 
abundances with open symbols. In the dI sample, NGC~5264 is once more an outlier. Rejecting 
this point, a geometric linear fit to the rest of the sample (28 dIs) gives the following 
relation: 
\begin{equation}
12+\log(\rm{O/H}) = (4.91 \pm 0.46) + (0.35 \pm 0.05) \log(M_{bary})
\end{equation}
\noindent
This relation is very close to the fit found by \citet{lee03a} which was derived from 
$B$ photometry for 21 dIs in the field. We plot our fit with a solid line and the fit 
of \citet{lee03a} with a dotted line. 
The close agreement suggests that the two-component decomposition scheme of 
\citep{lee03a} was effective for deriving stellar masses. Most BCDs in 
Figure~\ref{met_totalmass} appear to lie at lower baryonic masses compared to dIs at 
a given oxygen abundance, possibly as a consequence of how BCDs are selected. It is 
possible that larger galaxies with comparable bursts are simply not classified as 
BCDs! 

\subsection{Metallicity and Gas Fraction}

The closed box model of chemical evolution predicts a linear correlation between 
$n$(O)$/n$(H) and the logarithm of the inverse gas fraction $1/\mu$, where 
\begin{equation}
\mu = \frac{M_{gas}} {M_{gas}+M_{*}}
\end{equation}
The slope of the relation conveys the yield of oxygen. Judging stellar masses from 
BV photometry, \citet{lee03a} concluded that field dIs obey the closed box model, implying 
that their evolution has been isolated. Here, we study the correlation obtained for 
both dIs and BCDs using NIR sech luminosities to assess stellar masses. In principle, 
our masses should be more reliable, as they are less sensitive to recent star formation. 

\begin{figure}
\epsscale{1.0}
\plotone{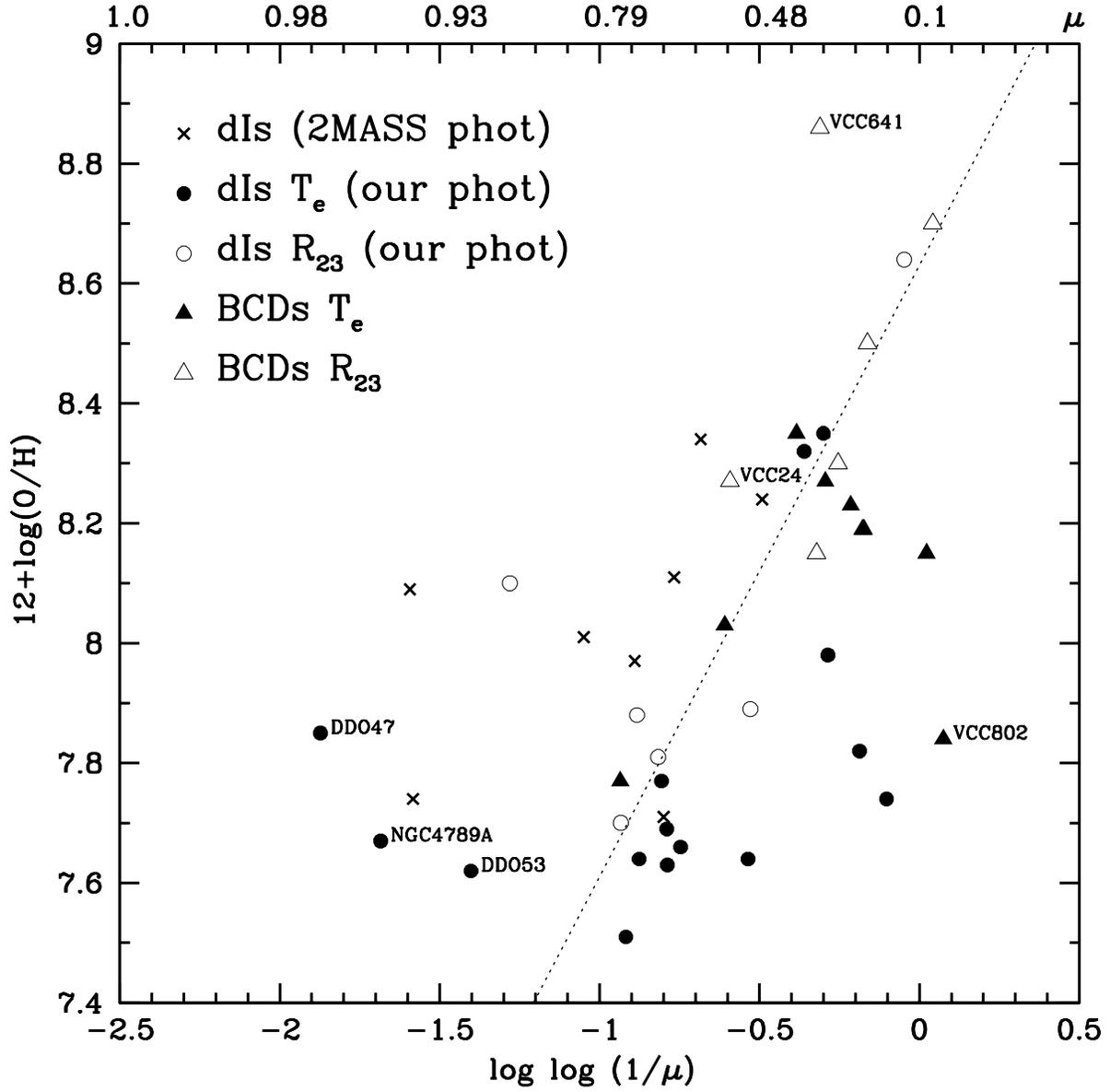}
\caption{
\label{miumiuz}
Correlation between oxygen abundance and the gas fraction, $\mu$. dIs whose stellar 
masses were derived from sech magnitudes in $K_s$ (Paper I) are plotted with circles, 
and while those derived from 2MASS photometry are plotted with crosses. 
BCDs, all of whose stellar masses were derived from sech 
magnitudes in $K_s$, are represented with triangles. $T_e$-based abundances are plotted 
as solid symbols and crosses, and $R_{23}$-based abundances as open symbols. 
The linear fit derived by \citet{lee03a} for his field dI sample is 
plotted with a dotted line. The gas fraction increases towards the left. 
Although there is a lot of scatter, most dIs and BCDs appear to follow the closed box model. 
}
\end{figure}

Figure~\ref{miumiuz} plots $\log(\rm{O/H})$ versus $\log(\log(1/\mu))$, for which the 
closed box model predicts a linear correlation with a slope of unity and an intercept 
equal to the yield. Gas fraction $\mu$ increases to the left, according to the labels 
plotted along the top horizontal axis. 
We represent dIs with photometry from the present work with circles, and dIs with 
2MASS photometry with crosses. We plot the fit of \citet{lee03a} with a dotted line. 
Most dIs from our sample lie close to the fit of Lee. 
We overlay in Figure~\ref{miumiuz} triangles representing our BCD sample. 
Most BCDs appear close to the fit of Lee, too, with about three outliers. 
Note that gas has not been detected in VCC 802, so its abscissa is a lower limit. 
One posibility is that VCC~802 has been stripped due to the hot intracluster medium since 
it is projected near the M86/M84 clump, which is a region of enhanced density \citep{boe94}. 
If dIs follow the closed box model, as the data suggests, then the proximity of BCDs to 
the dI locus suggests that bursts in present-day BCDs do not have sufficient strength 
to blow out neutral gas from the host galaxies. The fact that several dIs and at least 
one BCD do show HI deficiencies in dense environments (\citealp{lee03a}; this paper) 
indicates that external processes are more likely to remove gas from dwarfs. 
In summary, we concur with Lee that the evolution of field dIs has not been affected 
much by flows. We also conclude that BCDs have evolved similarly to dIs.

\section{Implications for Evolution}
\label{evolution}

A few evolutionary scenarios that involve dIs and BCDs have been proposed over the years,
most without firm conclusions (e.g., \citealp{pap96b}). Some links have emerged using stellar
masses derived from data in the visible, although with large scatter. We summarize below our
main results.

For both dIs and BCDs, metallicity correlates with stellar mass, gas mass,
and baryonic mass, in the sense that more massive systems are more enriched in metals.
Our most important discovery is that the relationship between the metallicity and the gas
fraction for BCDs follows that for dIs. It means that BCDs and dIs have evolved similarly.
Using NIR photometry, we confirm the conclusion of \citet{lee03a} that the evolution of
field dIs has not been noticeably influenced by gas flows. By implication, the same
conclusion must be reached for BCDs. Thus, star formation bursts in BCDs are unlikely
to be caused by the transfer of material from elsewhere.

\section{Conclusions}
\label{conclusions}

Oxygen abundances have been derived for four blue compact dwarf galaxies in the
Virgo Cluster observed in a queue run at Gemini-North in 2003.
Drawing upon the NIR properties studied in Papers I and II and oxygen abundances 
from our work and the literature, we studied the correlations between the metallicity 
and the stellar mass, gas mass, baryonic, mass, and the gas fraction for both dIs
and BCDs.
Metallicity correlates with all four parameters, in the sense that more massive 
systems contain more metals. The oxygen abundance correlates well with the luminosity 
in $K_s$, for both dIs and BCDs. 
Using NIR sech magnitudes to measure luminosity, the $L_K-Z$ relations for dIs and 
BCDs are statistically indistinguishable. 
dIs and BCDs appear to share a common relationship between metallicity and the gas 
fraction, namely one which is consistent with the closed-box evolution. Thus, BCDs appear to 
be dIs with enhanced star formation activity. The origin of the enhancement remains
unknown, but it is unlikely to be caused by flows.

\acknowledgments

We thank the Gemini-North time allocation committees for granting us the opportunity to observe.
MLM thanks the Natural Sciences and Engineering Research Council of Canada for its continuing 
support. MGR acknowledges financial support from CONACyT grant 43121 and UNAM DGAPA grants 
IN108406-2, IN108506-2, and IN112103.
OV thanks Dr. Henry Lee for providing up-to-date references regarding metallicities.
Special acknowledgement is due to Dr. Jose Manuel Vilchez who served as the external
examiner for the PhD thesis of OV, providing interesting discussions.
Thanks are due to Robin Fingerhut for providing updated distances of dwarfs in the Local Volume.
For our data reductions, we used the GEMINI package written at the Gemini Observatories and IRAF,
which is maintained by the National Optical Astronomy Observatories, operated by the
Association of Universities for Research in Astronomy, Inc., under cooperative agreement
with the National Science Foundation.
This research has made use of the GOLDMine Database in Milano and the NASA/IPAC
Extragalactic Database (NED) which is operated by the JPL, CALTECH, under contract with NASA.
Special acknowledgement is due to an anonymous referee who provided discussions which allowed 
us to improve the manuscript.

%
%

\clearpage
\begin{deluxetable}{lrrrcrc}
\tablewidth{0pt}
\tabletypesize{\scriptsize}
\tablecaption{Gemini-North observing log \label{tbl-1}}
\tablehead{
\colhead{Galaxy} &
\colhead{$\alpha$} &
\colhead{$\delta$} &
\colhead{$PA$} &
\colhead{Date} &
\colhead{$\lambda_c$} &
\colhead{Exposure} \\
\colhead{} &
\colhead{(h m s)} &
\colhead{($\deg$ $\arcmin$ $\arcsec$)} &
\colhead{(deg)} &
\colhead{(UT)} &
\colhead{(\AA)} &
\colhead{(no.exp $\times$ s)} \\
\colhead{(1)} &
\colhead{(2)} &
\colhead{(3)} &
\colhead{(4)} &
\colhead{(5)} &
\colhead{(6)} &
\colhead{(7)}
}
\startdata
VCC 24   & 12:10:35.7 & +11:45:38 & 147 & Mar 1, 2003 & 4680 & $3\times600$  \\
...      & ...        & ...       & 147 & Mar 1, 2003 & 5600 & $3\times200$  \\

VCC 459  & 12:21:12.1 & +17:38:21 & 111 & Mar 1, 2003 & 4680 & $3\times600$  \\
...      & ...        & ...       & 111 & Mar 1, 2003 & 5600 & $3\times200$  \\

VCC 641  & 12:23:28.6 & +05:48:59 &  15 & Mar 1, 2003 & 4680 & $3\times600$  \\
...      & ...        & ...       &  15 & Mar 1, 2003 & 5600 & $3\times200$  \\

VCC 2033 & 12:46:04.5 & +08:28:33 &  14 & Mar 2, 2003 & 4680 & $3\times600$  \\
...      & ...        & ...       &  14 & Mar 2, 2003 & 5600 & $3\times200$  \\
\enddata
\tablecomments{
(1) Designation of galaxy in Virgo Cluster Catalog (VCC)
(2) Right Ascension (J2000) for the centre of the slit
(3) Declination (J2000) for the centre of the slit
(4) Slit position angle (E of N)
(5) Observing date
(6) Central wavelength
(7) Exposure time
}
\end{deluxetable}

\begin{deluxetable}{lrrrrrrrr}
\tablewidth{0pt}
\tabletypesize{\scriptsize}
\tablecaption{Observed and corrected line ratios \label{VCC_lines}}
\tablehead{
  &
  \multicolumn{2}{c}{VCC 24 \tablenotemark{a}} &
  \multicolumn{2}{c}{VCC 459 \tablenotemark{a}} &
  \multicolumn{2}{c}{VCC 641 \tablenotemark{b}} &   
  \multicolumn{2}{c}{VCC 2033 \tablenotemark{a}} \\
  \\
  \cline{1-9}
  \multicolumn{1}{c}{Identification (\AA)} &
  \multicolumn{1}{c}{$F$} & \multicolumn{1}{c}{$I$} &
  \multicolumn{1}{c}{$F$} & \multicolumn{1}{c}{$I$} &
  \multicolumn{1}{c}{$F$} & \multicolumn{1}{c}{$I$} &
  \multicolumn{1}{c}{$F$} & \multicolumn{1}{c}{$I$}
  }
\startdata
$\rm{[O~II]}$ 3727 & 334.8$\pm$16.1 & 344.8$\pm$16.8 & 212.4$\pm$1.4 & 230.2$\pm$4.6 & 280.7$\pm$48.2 & 297.3$\pm$52.6 & 436.3$\pm$4.4 & 447.0$\pm$4.6 \\
$\rm{[H~12]}$ & \nodata & \nodata & 1.8$\pm$1.2 & 2.0$\pm$1.4 & \nodata & \nodata & \nodata & \nodata \\
$\rm{[H~11]}$ 3771 & \nodata & \nodata & 2.3$\pm$0.3 & 2.5$\pm$0.3 & \nodata & \nodata & \nodata & \nodata \\
$\rm{[H~10]}$ 3798 &	\nodata & \nodata & 3.5$\pm$0.3 & 3.8$\pm$0.4 & \nodata & \nodata & \nodata & \nodata \\
$\rm{[He~I]}$ 3820 &	\nodata & \nodata & 0.6$\pm$0.3 & 0.6$\pm$0.3 & \nodata & \nodata & \nodata & \nodata \\
$\rm{[H~9]}$ 3835 &	\nodata & \nodata & 4.3$\pm$0.3 & 4.7$\pm$0.4 & \nodata & \nodata & \nodata & \nodata \\
$\rm{[Ne~III]}$ 3869 & \nodata & \nodata & 28.3$\pm$0.6 & 30.3$\pm$1.0 & \nodata & \nodata & \nodata & \nodata \\
$\rm{He~I + H~8}$ 3889 & \nodata & \nodata & 15.7$\pm$0.5 & 16.9$\pm$0.7 & \nodata & \nodata & \nodata & \nodata \\
$\rm{[Ne~III]}$ 3967 & \nodata & \nodata & 20.2$\pm$0.4 & 21.5$\pm$0.6 & \nodata & \nodata & \nodata & \nodata \\
$\rm{[S~II]}$ 4069 &	\nodata & \nodata & 2.0$\pm$0.3 & 2.1$\pm$0.3 & \nodata & \nodata & \nodata & \nodata \\
$\rm{H\delta}$ 4101 & \nodata & \nodata & 22.3$\pm$0.3 & 23.6$\pm$0.6 & \nodata & \nodata & \nodata & \nodata \\
$\rm{H\gamma}$ 4340 & \nodata & \nodata & 45.3$\pm$0.4 & 47.1$\pm$0.9 & \nodata & \nodata & 47.0$\pm$2.0 & 47.5$\pm$2.1 \\
$\rm{[O~III]}$ 4363 & \nodata & \nodata & 2.8$\pm$0.3 & 3.0$\pm$0.4 & \nodata & \nodata & 2.1$\pm$0.9 & 2.1$\pm$0.9 \\
$\rm{[He~I]}$ 4472 & \nodata & \nodata & 3.5$\pm$0.1 & 3.6$\pm$1.5 & \nodata & \nodata & \nodata & \nodata \\
$\rm{H\beta}$ 4861 & 100.0$\pm$2.1 & 100.0$\pm$4.3 & 100.0$\pm$0.5 & 100.0$\pm$1.0 & \nodata & \nodata & 100.0$\pm$2.8 & 100.0$\pm$3.0 \\
$\rm{He~I}$ 4922 & \nodata & \nodata & 1.0$\pm$0.4 & 1.0$\pm$0.5 & \nodata & \nodata & \nodata & \nodata \\
$\rm{[O~III]}$ 4959 & 92.2$\pm$3.8 & 92.0$\pm$4.0 & 132.3$\pm$1.7 & 131.3$\pm$2.3 & 54.9$\pm$6.6 & 56.5$\pm$7.3 & 57.8$\pm$1.9 & 57.6$\pm$2.0 \\
$\rm{[O~III]}$ 5007 & 271.3$\pm$8.0 & 270.2$\pm$8.3 & 396.5$\pm$2.0 & 392.1$\pm$3.7 & 82.3$\pm$11.5 & 84.5$\pm$12.6 & 172.8$\pm$3.8 & 172.2$\pm$3.9 \\
$\rm{[He~I]}$ 5876 & \nodata & \nodata & 12.4$\pm$0.1 & 11.6$\pm$0.1 & 13.8 & 13.9$\pm$4.6$\pm$0.5 & 9.0$\pm$0.6 & 9.0$\pm$0.7 \\
$\rm{[O~I]}$ 6300 & \nodata & \nodata & 4.5$\pm$0.1 & 4.2$\pm$0.1 & \nodata & \nodata & \nodata & \nodata \\
$\rm{[S~III]}$ 6312 & \nodata & \nodata & 1.6$\pm$0.1 & 1.4$\pm$0.1 & \nodata & \nodata & \nodata & \nodata \\
$\rm{[O~I]}$ 6363 & \nodata & \nodata & 1.2$\pm$0.1 & 1.1$\pm$0.1 & \nodata & \nodata & \nodata & \nodata \\
$\rm{[N~II]}$ 6548 & 16.8$\pm$2.4 & 16.3$\pm$2.5 & 7.5$\pm$0.9 & 6.9$\pm$0.8 & \nodata & \nodata & 7.6$\pm$1.2 & 7.4$\pm$1.3 \\
$\rm{H\alpha}$ 6563 & 373.9$\pm$4.0 & 286.3$\pm$4.4 & 314.6$\pm$1.2 & 286.3$\pm$0.6 & 100.0$\pm$8.3 & 100.0$\pm$9.1 & 261.4$\pm$5.1 & 254.1$\pm$5.3 \\
$\rm{[N~II]}$ 6583 & 44.6$\pm$2.8 & 43.1$\pm$3.0 & 22.2$\pm$0.9 & 20.2$\pm$0.8 & 18.3$\pm$3.3 & 18.3$\pm$3.6 & 24.6$\pm$1.4 & 23.9$\pm$1.5 \\
$\rm{[He~I]}$ 6678 & 4.6$\pm$1.0 & 4.4$\pm$1.0 & 3.6$\pm$0.1 & 3.3$\pm$0.1 & \nodata & \nodata & 2.6$\pm$0.2 & 2.5$\pm$0.2 \\
$\rm{[S~II]}$ 6716 & 65.6$\pm$2.5 & 63.2$\pm$2.6 & 26.6$\pm$0.1 & 24.1$\pm$0.1 & 34.7$\pm$3.6 & 34.6$\pm$3.9 & 28.5$\pm$0.6 & 27.6$\pm$0.7 \\
$\rm{[S~II]}$ 6730 & 50.0$\pm$2.8 & 48.2$\pm$3.0 & 19.4$\pm$0.1 & 17.6$\pm$0.1 & 21.3$\pm$2.8 & 21.3$\pm$3.1 & 19.7$\pm$0.5 & 19.1$\pm$0.5
\enddata
\tablecomments{
$^a$ Flux ratios are reported with respect to H$\beta$~(H$\beta$=100) 
$^b$ Flux ratios are reported with respect to H$\alpha$~(H$\alpha$=100). 
Wavelengths are listed in \AA. $F$ is the observed flux ratio. $I$ is the flux ratio corrected for reddening. 
}
\end{deluxetable}

\clearpage
\begin{deluxetable}{ccccc}
\tabletypesize{\scriptsize}
\tablewidth{0pt}
\tablecaption{Derived properties for HII regions in Virgo BCDs \label{table_BCDs_props}}
\tablehead{
\colhead{Property} &
\colhead{VCC 24} &
\colhead{VCC 459} &
\colhead{VCC 641} &
\colhead{VCC 2033}
}
\startdata
$I(H\beta$) & (3.65$\pm$0.08) & (2.38$\pm$0.07) & \nodata & (3.62$\pm$0.05) \\
   (ergs s$^{-1}$ cm$^{-2}$) & $\times 10^{-15}$ & $\times 10^{-14}$ & \nodata & $\times 10^{-15}$ \\
$\tau_1$ & 0.032 & 0.087 & 0.028 & 0.026 \\
$W(H\beta$) (\AA) & 4.73$\pm$0.10 & 92.80$\pm$1.19 & \nodata & 11.10$\pm$0.17 \\
$n_e$ (cm$^{-3}$) & \nodata & 52 & \nodata & \nodata \\
log $R_{23}$ & 0.849 & 0.877 & 1.097 & 0.830 \\
log $O_{32}$ & 0.021 & 2.273 & -0.324 & -0.289 \\
$T_e$(O$^{+2}$) (K) & \nodata & 10500 & \nodata & 12500 \\
O$^+$/H$^+$    & \nodata & (6.9$\pm$1.4)$\times 10^{-5}$ & \nodata & (6.9$\pm$0.7)$\times 10^{-5}$ \\
O$^{+2}$/H$^+$ & \nodata & (1.2$\pm$0.3)$\times 10^{-4}$ & \nodata & (2.9$\pm$1.3)$\times 10^{-5}$ \\
$12+\log(\rm{O/H})$ $[T_e]$     & \nodata  & 8.27$\pm$0.09 & \nodata & 7.99$\pm$0.30 \\
$12+\log(\rm{O/H})$ $[R_{23}]$  & 8.58$\pm$0.20 & 8.09$\pm$0.20 & 8.86$\pm$0.20 & 8.31$\pm$0.20 \\ 
\enddata
\tablecomments{
I(H$\beta$) represents the H$\beta$ intensity, in ergs s$^{-1}$ cm$^{-2}$, corrected for underlying Balmer absorption and reddening;
$\tau_1$ is the optical depth of dust at 1 $\mu m$, assuming a Fitzpatrick (1999) reddening model and dust in the foreground located in the Milky Way, calculated using H$\alpha$/H$\beta$;
W(H$\beta$) is the emission equivalent width, in \AA, corrected for underlying Balmer absorption;
$n_e$ is the electron density;
$\log(R_{23})$ and $\log(O_{32})$ are the diagnostics for the bright-line method (\citealp{pag79}, \citealp{mcg94});
$T_e(\rm{O}^{+2})$ is the computed electron temperature in degrees Kelvin;
$12+\log(\rm{O/H})$ $[T_e]$ is the derived oxygen abundance in dex measured via $T_e$;
$12+\log(\rm{O/H})$ $[R_{23}]$ is the derived oxygen abundance in dex measured via $R_{23}$.
}
\end{deluxetable}

\clearpage
\begin{deluxetable}{lrrrrrcrrr}
\tablewidth{0pt}
\tabletypesize{\scriptsize}
\tablecaption{Properties of the dI sample \label{dIs_met}}
\tablehead{
\colhead{Galaxy} &
\colhead{DM} &
\colhead{$M_K$} &
\colhead{$\log F_{21}$} &
\colhead{$\log M_{gas}$} &
\colhead{$\log M_*$} &
\colhead{$\mu$} &
\colhead{12+log(O/H)} &
\colhead{Method} &
\colhead{Reference} \\
\colhead{(1)} &
\colhead{(2)} &
\colhead{(3)} &
\colhead{(4)} &
\colhead{(5)} &
\colhead{(6)} &
\colhead{(7)} &
\colhead{(8)} &
\colhead{(9)} &
\colhead{(10)}
}
\startdata
Cassiopeia 1  & 27.59 & -18.23 & 1.70 & 8.24 & 8.51 & \nodata & \nodata & \nodata  &\nodata \\
MB 1          & 27.59 & -17.56 & 0.91 & 7.45 & 8.25 & \nodata & \nodata & \nodata  &\nodata \\
UGCA 92       & 26.20 & -15.68 & 2.02 & 8.01 & 7.49 & 0.765 & 7.70 $\pm$ 0.20 & $R_{23}$ & HM95  \\
Orion dwarf   & 28.66 & -17.96 & 1.89 & 8.86 & 8.41 & 0.740 & 7.88 $\pm$ 0.20 & $R_{23}$ & LFM05 \\
DDO 47        & 28.39 & -15.04 & 1.88 & 8.74 & 7.24 & 0.970 & 7.85 $\pm$ 0.04 & $T_e$    & SKH89 \\
UGC 4115      & 28.53 & -16.60 & 1.32 & 8.24 & 7.86 & 0.704 & 7.81 $\pm$ 0.20 & $R_{23}$ & LFM05 \\
DDO 53        & 27.58 & -13.59 & 1.14 & 7.68 & 6.66 & 0.913 & 7.62 $\pm$ 0.05 & $T_e$    & SKH89 \\
UGC 4483      & 27.37 & -14.67 & 1.13 & 7.58 & 7.09 & 0.757 & 7.51 $\pm$ 0.03 & $T_e$    & STK94 \\
UGC 4998      & 29.95 & -18.53 & \nodata & \nodata & \nodata & \nodata & \nodata & \nodata & \nodata \\
UGC 5423      & 28.11 & -16.17 & 0.58 & 7.33 & 7.69 & 0.304 & 7.98 $\pm$ 0.10 & $T_e$    & MH96  \\
UGC 5692      & 27.83 & -17.62 & \nodata & \nodata & \nodata & \nodata & \nodata & \nodata & \nodata \\
UGC 5848      & 29.70 & -18.11 & \nodata & \nodata & \nodata & \nodata & \nodata & \nodata & \nodata \\
UGC 5979      & 30.55 & -18.49 & \nodata & \nodata & \nodata & \nodata & \nodata & \nodata & \nodata \\
UGC 6456      & 28.03 & -15.50 & 1.15 & 7.87 & 7.42 & 0.736 & 7.64 $\pm$ 0.10 & $T_e$    & TBD81 \\
Markarian 178 & 27.78 & -16.04 & 0.48 & 7.10 & 7.64 & 0.224 & 7.82 $\pm$ 0.06 & $T_e$    & GIT00 \\
NGC 3741      & 27.24 & -15.02 & 1.72 & 8.12 & 7.23 & 0.886 & 8.10 $\pm$ 0.20 & $R_{23}$ & GH89  \\
NGC 4163      & 27.30 & -16.34 & 0.98 & 7.41 & 7.76 & \nodata & \nodata & \nodata & \nodata \\
NGC 4190      & 27.20 & -16.47 & 1.37 & 7.76 & 7.81 & \nodata & \nodata & \nodata & \nodata \\
Markarian 209 & 28.37 & -15.67 & 1.00 & 7.85 & 7.49 & 0.698 & 7.77 $\pm$ 0.01 & $T_e$    & IT99  \\
NGC 4789A     & 27.80 & -15.63 & 2.16 & 8.79 & 7.47 & 0.953 & 7.67 $\pm$ 0.05 & $T_e$    & KS01  \\
GR 8          & 26.52 & -13.60 & 0.89 & 7.00 & 6.66 & 0.687 & 7.63 $\pm$ 0.10 & $T_e$    & MAM90 \\
DDO 167       & 27.93 & -14.56 & 0.66 & 7.34 & 7.05 & 0.662 & 7.66 $\pm$ 0.03 & $T_e$    & SKH89 \\
UGC 8508      & 26.87 & -15.43 & 1.15 & 7.40 & 7.39 & 0.506 & 7.89 $\pm$ 0.20 & $R_{23}$ & LFM05 \\
NGC 5264      & 28.11 & -18.64 & 1.09 & 7.84 & 8.68 & 0.127 & 8.64 $\pm$ 0.20 & $R_{23}$ & LGH03 \\
Holmberg IV   & 29.22 & -18.19 & 1.30 & 8.49 & 8.50 & \nodata & \nodata & \nodata & \nodata \\
DDO 187       & 26.78 & -14.33 & 1.08 & 7.30 & 6.95 & 0.688 & 7.69 $\pm$ 0.09 & $T_e$    & LMK03 \\
\\
VCC 428       & 30.62 & -15.78 &$-$0.20 & 7.55 & 7.53 & 0.512 & 7.64 $\pm$ 0.09 & $T_e$    & VP03  \\
VCC 1374      & 30.62 & -18.17 & 0.40 & 8.15 & 8.49 & 0.315 & 8.35 $\pm$ 0.14 & $T_e$    & VP03  \\
VCC 1699      & 30.62 & -18.90 & 0.79 & 8.54 & 8.78 & 0.366 & 8.32 $\pm$ 0.06 & $T_e$    & VP03  \\
VCC 1725      & 30.62 & -18.81 & 0.28 & 8.03 & 8.75 & 0.162 & 7.74 $\pm$ 0.14 & $T_e$    & VP03  \\
\\
NGC 55        & 25.94 & -19.69 & 3.43 & 9.31 & 9.10 & 0.621 & 8.34 $\pm$ 0.10 & $T_e$ & WS83  \\
NGC 1560      & 27.07 & -18.25 & 2.65 & 8.98 & 8.52 & 0.743 & $>$ 7.97        & $T_e$ & LMK03 \\
Holmberg II   & 27.50 & -18.72 & 2.56 & 9.07 & 8.71 & 0.694 & 7.71 $\pm$ 0.13 & $T_e$ & LMK03 \\
NGC 3109      & 25.72 & -16.46 & 3.22 & 9.01 & 7.81 & 0.942 & 7.74 $\pm$ 0.33 & $T_e$ & LMK03 \\
IC 2574       & 27.86 & -17.15 & 2.65 & 9.30 & 8.08 & 0.943 & 8.09 $\pm$ 0.07 & $T_e$ & MH96  \\
NGC 4214      & 28.25 & -20.34 & 2.51 & 9.32 & 9.36 & 0.476 & 8.24 $\pm$ 0.12 & $T_e$ & KS96  \\
NGC 5408      & 27.76 & -16.39 & 1.81 & 8.42 & 7.78 & 0.814 & 8.01 $\pm$ 0.02 & $T_e$ & SCV86 \\
NGC 6822      & 23.46 & -16.83 & 3.38 & 8.27 & 7.95 & 0.674 & 8.11 $\pm$ 0.11 & $T_e$ & L05
\enddata
\tablecomments{
(1) Galaxy name, ordered by right ascension. The first two groups list our field and Virgo observed samples, and the third group lists the the 2MASS sample 
(2) Distance modulus
(3) Absolute magnitude in $K_s$ (sech for our sample, total for 2MASS)
(4) Logarithm of the total HI flux in Jy km s$^{-1}$ \citep{kar04}
(5) Logarithm of the total gas mass in solar masses
(6) Logarithm of the stellar mass in solar masses
(7) Gas fraction
(8) Oxygen abundance and uncertainty
(9) Method of determining abundances: $T_e$-direct determination via [O III]$\lambda$4363 \citep{ost89}; $R_{23}$-determination via the bright line method \citep{mcg94}
(10) Reference for abundance: KS97: \citet{kob97}; HM95: \citet{hod95}; LFM05: Lee, Fingerhut and McCall, 2005 (in progress);
SKH89: \citet{ski89}; STK94: \citet{ski94}; MH96: \citet{mil96}; TBD81: \citet{tul81};
GIT00: \citet{gus00}; M97: \citet{m97}; GH89: \citet{gal89}; IT89: \citet{izo99};
KS01: \citet{ken01}; MAM90: \citet{mol90}; LGH03: \citet{lee03c}; LMK03: \citet{lee03a};
WS83: \citet{web83}; KS96: \citet{kob96}; SKV86: \citet{sta86}; L05: Lee, H., Skillman, E. D., \& Venn, K. A., 2005, private communication; 
VP03: \citet{vil03}
}
\end{deluxetable}

%
%

\clearpage
\begin{deluxetable}{lrrrrrrcrrr}
\tablewidth{0pt}
\tabletypesize{\scriptsize}
\tablecaption{Properties of the BCD sample \label{BCDs_met}}
\tablehead{
\colhead{Galaxy} &
\colhead{DM} &
\colhead{$M_K$} &
\colhead{$\log F_{21}$} &
\colhead{$\log M_{gas}$} &
\colhead{$\log M_*$} &
\colhead{$\mu$} &
\colhead{12+log(O/H)} &
\colhead{Method} &
\colhead{Reference} \\
\colhead{(1)} &
\colhead{(2)} &
\colhead{(3)} &
\colhead{(4)} &
\colhead{(5)} &
\colhead{(6)} &
\colhead{(7)} &
\colhead{(8)} &
\colhead{(9)} &
\colhead{(10)}
}
\startdata
VCC 24   & 30.62 & -17.54 & 0.58 & 8.33 & 8.24 & 0.553 & 8.27 $\pm$ 0.20 & $R_{23}$ & V05   \\
VCC 144  & 30.62 & -17.73 & 0.36 & 8.11 & 8.31 & 0.387 & 8.35 $\pm$ 0.07 & $T_e$    & VP03  \\
VCC 213  & 30.62 & -19.42 & 0.17 & 7.92 & 8.99 & 0.079 & 8.70 $\pm$ 0.20 & $R_{23}$ & VP03  \\
VCC 324  & 30.62 & -18.70 & 0.36 & 8.11 & 8.70 & 0.205 & 8.50 $\pm$ 0.10 & $R_{23}$ & VP03  \\
VCC 334  & 30.62 & -17.38 & 0.12 & 7.87 & 8.17 & 0.335 & 8.15 $\pm$ 0.05 & $R_{23}$ & VP03  \\
VCC 459  & 30.62 & -18.17 & 0.39 & 8.14 & 8.49 & 0.310 & 8.27 $\pm$ 0.09 & $T_e$    & V05   \\
VCC 641  & 30.62 & -17.25 & 0.05 & 7.81 & 8.12 & 0.325 & 8.86 $\pm$ 0.20 & $R_{23}$ & V05   \\
VCC 802  & 30.62 & -16.40 & $<-$1.13 & 6.62 & 7.78 & 0.065 & 7.84 $\pm$ 0.15 & $T_e$ & VP03  \\
VCC 848  & 30.62 & -17.91 & 0.75 & 8.51 & 8.39 & 0.568 & 8.03 $\pm$ 0.16 & $T_e$    & VP03  \\
VCC 1313 & 30.62 & -15.07 & 0.01 & 7.76 & 7.25 & 0.765 & 7.77 $\pm$ 0.06 & $T_e$    & VP03  \\
VCC 1437 & 30.62 & -18.37 & 0.40 & 8.15 & 8.57 & 0.277 & 8.30 $\pm$ 0.20 & $R_{23}$ & VP03  \\
VCC 2033 & 30.62 & -17.91 &$-$0.38 & 7.37 & 8.39 & 0.088 & 8.15 $\pm$ 0.15 & $T_e+R_{23}$ & V05 \\
\\
NGC 1569 & 26.37 & -18.19 & 1.87 & 8.05 & 8.49 & 0.268 & 8.19 $\pm$ 0.04 & $T_e$    & KS97  \\ 
NGC 3738 & 28.27 & -18.59 & 1.34 & 8.28 & 8.64 & 0.305 & 8.23 $\pm$ 0.01 & $T_e$    & M97   \\ 
IC 10    & 24.35 & -18.90 & 2.98 & 8.36 & 8.78 & 0.276 & 8.19 $\pm$ 0.14 & $T_e$    & LMK03 \\ 
\enddata
\tablecomments{
(1) Galaxy name, ordered by number. The first group represents our Virgo observed sample, and the second our BCD field sample 
(2) Distance modulus
(3) Sech absolute magnitude in $K_s$
(4) Logarithm of the total HI flux in Jy km s$^{-1}$ based on \citet{gav05}
(5) Logarithm of the total gas mass in solar masses
(6) Logarithm of the stellar mass in solar masses
(7) Gas fraction
(8) Oxygen abundance and uncertainty
(9) Method of determining abundances: $T_e$-direct determination via [O III]$\lambda$4363 \citep{ost89}; $R_{23}$-determination via the bright line method \citep{mcg94}
(10) Reference for abundances: V05: this work; VP03: \citet{vil03}; KS97: \citet{kob97}; M97: \citet{m97}
}
\end{deluxetable}

\clearpage
\begin{deluxetable}{lrrrrr}
\tablewidth{0pt}
\tabletypesize{\scriptsize}
\tablecaption{Luminosity-metallicity relations for star-forming galaxies \label{LZ_comp}}
\tablehead{
\colhead{Author} &
\colhead{Band} &
\colhead{Sample} &
\colhead{$L-Z$ Slope} &
\colhead{$L-Z$ Intercept} &
\colhead{rms} \\
\colhead{(1)} &
\colhead{(2)} &
\colhead{(3)} &
\colhead{(4)} &
\colhead{(5)} &
\colhead{(6)}
}
\startdata
SKH89    & $B$ & 19 dIs & -0.153             & 5.500             & 0.16  \\
RM95     & $B$ & 12 dIs & -0.147 $\pm$ 0.029 & 5.670 $\pm$ 0.480 & 0.09  \\
LMK03    & $B$ & 22 dIs & -0.153 $\pm$ 0.025 & 5.590 $\pm$ 0.540 & 0.17  \\
ZH06     & $B$ & 21 dIs & -0.149 $\pm$ 0.011 & 5.650 $\pm$ 0.170 & 0.15  \\ 

LMC04    & $B$ & $\times$100 MP SFGs from 2dFGRS & -0.200 $\pm$ 0.020 & 4.290 $\pm$ 0.420 & $>$0.3  \\
LMC04    & $B$ & 6387 SFGs from 2dFGRS           & -0.223 $\pm$ 0.004 & 4.070 $\pm$ 0.090 & 0.32 \\

THK04    & $B$ & $\sim$53000 SFGs from SDSS & -0.185 $\pm$ 0.001 & 5.238 $\pm$ 0.018 & $>$0.3  \\

SKL05    & $B$ & 24 dwarf BCGs $M_B>-18$ & -0.139 $\pm$ 0.011 & 5.800 $\pm$ 0.170 & $>$0.2  \\
SKL05    & $B$ & 48 BCGs $M_B<-18$       & -0.079 $\pm$ 0.018 & 6.930 $\pm$ 0.370 & $>$0.2  \\

SLM05    & $B$ & 765 SFGs from KISS & -0.222 $\pm$ 0.003 & 4.180 $\pm$ 0.060 & 0.25  \\
SLM05    & $J$ & 420 SFGs from KISS & -0.200 $\pm$ 0.004 & 4.130 $\pm$ 0.090 & 0.22  \\
SLM05    & $K$ & 370 SFGs from KISS & -0.195 $\pm$ 0.004 & 4.030 $\pm$ 0.010 & 0.23  \\

O06      & $K$ & 29 dwarfs from Paper I and 2MASS & -0.140 $\pm$ 0.020 & 5.550 $\pm$ 0.260 & 0.15 \\

V05      & $J$ & 25 dIs  & -0.140 $\pm$ 0.014 & 5.692 $\pm$ 0.214 & 0.11  \\
V05      & $K$ & 25 dIs  & -0.141 $\pm$ 0.015 & 5.581 $\pm$ 0.244 & 0.10  \\

V05      & $J$ & 14 BCDs & -0.216 $\pm$ 0.027 & 4.530 $\pm$ 0.473 & 0.12  \\
V05      & $K$ & 14 BCDs & -0.224 $\pm$ 0.030 & 4.212 $\pm$ 0.537 & 0.11  \\
\enddata
\tablecomments{
(1) Reference: SKH89: \citet{ski89}; RM95: \citet{ric95}; LMK03: \citet{lee03a}; ZH06: \citet{van06b}; 
    LMC04: \citet{lam04}; THK04: \citet{tre04}; SKL05: \citet{shi05}; SLM05: \citet{sal05}; 
    O06: \citet{oli06}; V05: Vaduvescu 2005 (this work);
(2) Observing Band (for the absolute magnitude);
(3) Sample: SFGs: Star-Forming Galaxies; BCGs: Blue Compact Galaxies; MP: metal-poor ($12+\log(\rm{O/H})\la8.3$);
(4) $L-Z$ relation slope;
(5) $L-Z$ relation intercept;
(6) root mean square error in log(O/H).
}
\end{deluxetable}

%
%

\clearpage

\clearpage

\clearpage

\clearpage

\clearpage

\clearpage

\clearpage

\clearpage

\clearpage

\clearpage

\end{document}